\documentclass[prd,aps,twocolumn,a4paper,floatfix,nofootinbib,longbibliography]{revtex4-1}

\usepackage{bm}
\usepackage{lipsum}
\usepackage{graphicx,psfrag}
\usepackage{enumerate}
\usepackage{mathrsfs}
\usepackage{amsmath,amsfonts,amssymb}
\usepackage{enumerate}
\usepackage{comment}
\usepackage{hyperref}
\usepackage{float}
\usepackage{ulem}
\usepackage{bbm}

\usepackage{color}

\newcommand{\p}{\partial}
\newcommand{\ul}{\underline}
\newcommand{\Lie}{\mathcal{L}}

\def\perpN{{}^{\textrm{\tiny{(N)}}}\!\!\!\perp}
\def\DN{{}^{\textrm{\tiny{(N)}}}D}
\def\GammaN{{}^{\textrm{\tiny{(N)}}}\Gamma}
\def\tinyN{{}^{\textrm{\tiny{(N)}}}}
\def\tinythree{{}^{\textrm{\tiny{(3)}}}}

\def\perpdsq{{}^{\mathbbmss{q}}\!\!\!\perp}

\begin{document}

\title{An Implementation of DF-GHG with Application to Spherical Black
  Hole Excision}

\author{Maitraya K \surname{Bhattacharyya}${}^{1,2}$}
\author{David \surname{Hilditch}${}^{3}$}
\author{K Rajesh \surname{Nayak}${}^{1,2}$}
\author{Sarah Renkhoff${}^{5}$}
\author{Hannes R \surname{R{\"u}ter}${}^{4}$}
\author{Bernd \surname{Br{\"u}gmann}${}^{5}$}
  
\affiliation{${}^{1}$Indian Institute of Science Education and
  Research Kolkata, Mohanpur 741246, India \\ ${}^{2}$Center of
  Excellence in Space Sciences India, Mohanpur 741246, India\\
  ${}^{3}$Centro de Astrof\'{\i}sica e Gravita\c c\~ao  - CENTRA,
  Departamento de F\'{\i}sica, Instituto Superior T\'ecnico - IST,
  Universidade de Lisboa - UL,
  Av. Rovisco Pais 1, 1049-001 Lisboa, Portugal\\
  ${}^{4}$Max Planck Institute for Gravitational Physics (Albert
  Einstein Institute), 14476 Potsdam-Golm, Germany\\
  ${}^{5}$Theoretical Physics Institute, University of Jena, 07743
  Jena, Germany}
\date{\today}

\begin{abstract}
We present an implementation of the dual foliation generalized
harmonic gauge (DF-GHG) formulation within the pseudospectral code
\texttt{bamps}. The formalism promises to give greater freedom in the
choice of coordinates that can be used in numerical relativity. As a
specific application we focus here on the treatment of black holes in
spherical symmetry. Existing approaches to black hole excision in
numerical relativity are susceptible to failure if the boundary fails
to remain outflow. We present a method, called DF-excision, to avoid
this failure. Our approach relies on carefully choosing coordinates in
which the coordinate lightspeeds are under strict control. These
coordinates are then combined with the DF-GHG formulation. After
performing a set of validation tests in a simple setting, we study the
accretion of large pulses of scalar field matter on to a spherical
black hole. We compare the results of DF-excision with a naive
setup. DF-excision proves reliable even when the previous approach
fails.
\end{abstract}

\maketitle

\section{Introduction}

Free-evolution formulations of GR for numerical relativity (NR) are
built with a number of requirements in mind. Foremost in this list is
that the specific PDE problem to be solved must be well-posed. The
easiest way to guarantee well-posedness of the initial value problem
is to try and render the equations hyperbolic so that textbook
theorems may be applied. This, in turn, requires a choice of
gauge. Considering the popular harmonic gauge choice~$\Box
X^{\ul{\alpha}}=0$ we see already that that such a choice requires a
choice of coordinates. But, in case we already have a choice of
coordinates in mind that do not satisfy this condition, the latter may
be problematic. It turns out that what is {\it really} required for
hyperbolicity is a sensible choice of tensor basis. If, with this
tensor basis fixed we change coordinates it turns out that in many
cases the equations remain hyperbolic. This strategy is regularly used
within the SpEC numerical relativity code~\cite{SchPfeLin06,SpEC} to
treat compact binary systems with coordinates that are approximately
corotating with the system, but always with a single foliation of
spacetime by a time coordinate~$T$. To overcome this restriction one
may turn to the dual-foliation (DF) formalism which, as first
presented in~\cite{Hil15}, allows us to employ a tensor basis
associated with coordinates~$X^{\ul{\alpha}}=(T,X^{\ul{i}})$ whilst
actually {\it working} in coordinates~$x^\alpha=(t,x^i)$. The DF
formalism has been used in a number of places in the
literature~\cite{HilRui16,HilHarBug16,SchHilBug17,HilSch18,GasHil18,
  DuaHil19,GasGauHil19,GauVanHil21} for mathematical analysis and is
under active investigation for the treatment of future null infinity.

In this paper, we present the first implementation of the
dual-foliation generalized harmonic gauge (DF-GHG) formulation of GR,
which was made in our pseudospectral
code~\texttt{bamps}~\cite{Bru11,HilWeyBru15}. In performing the
implementation we have made a number of validation tests, a few of
which are presented below. But to try and demonstrate the potential of
the formalism, we concentrate primarily on the specific use case of
black hole excision. The numerical binary black hole
breakthrough~\cite{Pre05,BakCenCho05,CamLouMar05} rests, loosely
speaking, on the backs of two different approaches for treating the
strong-field region, black hole excision and the moving-puncture
method. Each has strengths and weaknesses. Excision, as suggested by
Unruh to Thornburg~\cite{Tho87} and developed by many authors,
see~\cite{AlcBru00,SeiSue92,AnnDauMas95,CooHuqKla98,Tho99,ShoSmiSpe03,
  CalLehReu03,Pre04,SpeKelLag05} for a selection, relies on the idea
that nothing can escape from the black hole region, so it should be
possible to simply remove that region from the computational domain
without affecting the domain of outer communication whatsoever. This
has the advantage that the most violent spacetime region is not
treated, and the remaining solution may be reasonably expected to be
smooth. There are, however, two important requirements to
overcome. First, the intuitive idea that {\it nothing can escape}
needs to be encoded in a formal sense within the equations. This is
not trivial because if the excision boundary flaps around wildly and
fails to remain an outflow boundary, then we need to give boundary
conditions. Even in the Minkowski spacetime it is possible to
introduce an excision boundary that satisfies the first condition, by
simply taking a sphere and expanding it radially at the speed of
light. Secondly therefore, we must guarantee that the physical domain
is not discarded at the speed of light. For this we need to insure
that a small part of the black hole region stays within the
computational domain. Assuming that the apparent horizon remains
inside spatial slices of the event horizon, this could be done by
making sure that the apparent horizon remains on the grid. A final,
less fundamental, but nevertheless desirable property is to control
the coordinate position of the apparent horizon within the domain.

Within the SpEC code these necessary conditions for excision are
enforced by choosing spatial coordinates~$x^i$ with a control
system~\cite{HemSchKid13} that monitors the position of the apparent
horizon and drives the coordinates in a desirable direction. This
approach is very effective in practice, but as far as we are aware is
not guaranteed never to fail, even in spherical symmetry. It also has
the technical disadvantage that because the notion of apparent horizon
is quasilocal, it is not obvious that textbook well-posedness results
can be applied directly. In this paper we use the DF-GHG formulation
with coordinates carefully chosen for excision. Although we work in
the spherical setting we believe that it may eventually be possible to
use the key ingredients of our method in a more general context,
subsuming our coordinate choice within the control system
setup. Unsurprisingly the core point of our coordinate choice is the
use of an area-locking radial coordinate which, combined with insights
from the dynamical horizons framework~\cite{Hay94,AshKri03} guarantees
the first two properties mentioned above. In the near-future the
development presented here also has the important use that it will
allow us to generalize our earlier perturbative
work~\cite{BhaHilNay20} on the spherical scalar field within a fixed
Schwarzschild background to treat perturbations robustly in the fully
nonlinear setting, which the naive approach of~\cite{HilWeyBru17} was
incapable of. These results will be reported upon elsewhere.

We begin in Section~\ref{Section:DF-GHG} with an overview of the
DF-GHG formulation. In Section~\ref{Section:DFJac} we then describe,
at the continuum level, each of the coordinate choices that we test in
our implementation. In Section~\ref{Section:Code-Setup} we give a
brief overview overview of the~\texttt{bamps} code, before presenting
our results in Section~\ref{Section:Numerics}. Finally we conclude in
Section~\ref{Section:Conclusions}. Geometric units are used
throughout.

\section{The dual foliation formulation}\label{Section:DF-GHG}

\begin{figure}[t] 
  \centering \includegraphics[width=0.5\textwidth]{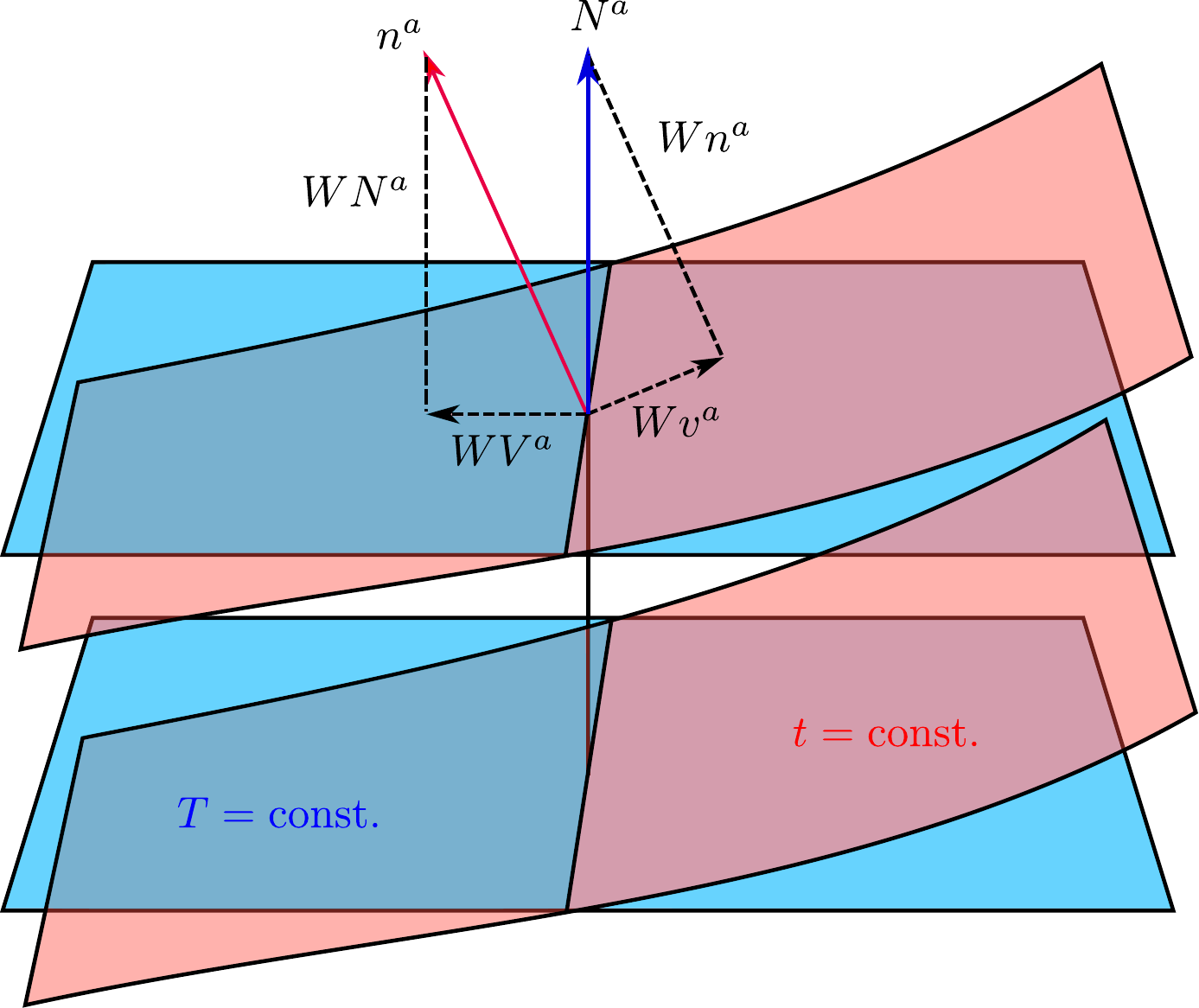}
  \caption{The DF approach: A spacetime with two different slicings
    and two coordinate systems, the upper case
    coordinates~$(T,X^{\ul{i}})$ and the lower case
    coordinates~$(t,x^{i})$. $N^a$ and~$n^a$ denote the timelike unit
    normal vectors for the two slices, the inner product of which is
    the Lorentz factor~$W = - (N^a n_a)$. $V^a = \frac{1}{W}
    \perpN^{b}_{\ a} n_b$ and~$v^a = \frac{1}{W} \perp^{b}_{\ a} N_b$
    denote the two boost vectors.}
  \label{fig:DFfoliation}
\end{figure}

In this section, we provide a brief summary of the dual foliation (DF)
formalism. Readers interested in a more detailed approach to the topic
may look at~\cite{Hil15} and~\cite{HilHarBug16}. The principal idea
behind the DF approach is to consider two coordinate systems defined
in the same region of spacetime~$x^{\mu} = (t,x^{i})$
and~$X^{\ul{\mu}} = (T,X^{\ul{i}})$, hereby referred to as the lower
case and upper case coordinates respectively. They are shown in
Fig.~\ref{fig:DFfoliation}. As a matter of convention, Greek indices
go over space and time, Latin indices~$a,b,c,d,e$ stand for abstract
indices, whereas~$i,j,k,l,m,p$ represent spatial components in
the~$x^{\mu}$ basis, and when underlined stand for spatial components
in the~$X^{\ul{\mu}}$ basis.

The two time coordinates~$t$ and~$T$ define two foliations of
spacetime, the lower case and upper case foliation respectively. In
practical applications of the DF formalism we aim to exploit good
properties of each coordinate system. As mentioned in the
introduction, in our specific setting, this will mean choosing the
upper case coordinates (and their associated tensor basis) to be
generalized harmonic~$\Box X^{\ul{\alpha}}=H^{\ul{\alpha}}$, which is
then used to guarantee symmetric hyperbolicity of the field equations
we solve. In later sections we will see that we can then choose the
lower case coordinates~$x^\mu$ in a variety of ways, including choices
that are useful for black hole excision.

In the lower case foliation, we can define the lapse, normal vector,
time vector, projection operator and shift vector as
\begin{align}
\alpha = (- \nabla_a t \nabla^a t)^{-\frac{1}{2}}, &&
n^a = - \alpha \nabla^a t, \nonumber \\
t^a \nabla_a t \equiv 1, && \perp^a_{\ b} = \delta^a_{\ b}
+ n^a n_b, \nonumber \\
\beta_a = \perp^b_{\ a} t_b, && \beta^i
= - \alpha n^{a} \nabla_a x^i.
\end{align}
Similar quantities may be defined in the upper case foliation
\begin{align}
A = (- \nabla_a T \nabla^a T)^{-\frac{1}{2}}, && N^a
= - A \nabla^a T, \nonumber \\
T^a \nabla_a T \equiv 1, && \perpN^a_{\ b}
= \delta^a_{\ b} + N^a N_b, \nonumber \\
B_a = \perpN^b_{\ a} T_b, && B^{\ul{i}}
= - A N^{a} \nabla_a X^{\ul{i}}.
\end{align}
The projection operator~$\perp^{a}_{\ b}$ with two indices downstairs
is the natural induced metric~$\gamma_{ab}$ on the lower case
foliation and a similar result follows for the upper case foliation
where the naturally induced metric is denoted
by~${}^{\textrm{\tiny{(N)}}}\gamma_{ab}$. The covariant derivative
associated with~$\gamma_{ab}$ is denoted by~$D$ and the corresponding
connection is denoted by~$\Gamma$. For the upper case spatial
metric~${}^{\textrm{\tiny{(N)}}}\gamma_{ab}$, the associated covariant
derivative is~$\DN$ and the corresponding connection is~$\GammaN$.

The relationship between the upper case and the lower case unit normal
vector is given by
\begin{align}
N^a = W (n^a + v^a), && n^a = W (N^a + V^a),
\end{align}
where~$W$ is called the Lorentz factor and is defined as
\begin{align}
W = - (N^a n_a) = \frac{1}{\sqrt{1 - v_i v^i}} =
\frac{1}{\sqrt{1 - V_{\ul{i}} V^{\ul{i}}}},
\end{align}
and~$V_a$ and~$v_a$ are the upper case and lower case boost vectors
defined as
\begin{align}
v_a = \frac{1}{W} \perp^{b}_{\ a} N_b, &&
V_a = \frac{1}{W} \perpN^{b}_{\ a} n_b.
\end{align}
The Jacobian matrix, defined as~$J^{\ul{\mu}}_{\ \mu} \equiv \p
X^{\ul{\mu}}/\p x^{\mu}$ can be decomposed in the~$3+1$ form
\begin{align}
n^{\alpha} J^{\ul{\alpha}}_{\ \alpha} N_{\ul{\alpha}} = - W, &&
n^{\alpha} J^{\ul{i}}_{\ \alpha} \equiv \pi^{\ul{i}}, \nonumber \\
J^{\ul{\alpha}}_{\ i} N_{\ul{\alpha}} = W v_i, && J^{\ul{i}}_{\ i}
\equiv \phi^{\ul{i}}_{\ i}.
\end{align}
In matrix form, the Jacobian can be represented as
\begin{align} \label{eq:jac}
J = \begin{pmatrix}
A^{-1} W (\alpha - \beta^i v_i) & \alpha \pi^{\ul{i}}
+ \beta^i \phi^{\ul{i}}_{\ i} \\
- A^{-1} W v_i & \phi^{\ul{i}}_{\ i}
\end{pmatrix},
\end{align}
with the inverse being
\begin{align} \label{eq:inversejac}
J^{-1} = \begin{pmatrix}
\alpha^{-1} W (A - B^{\ul{i}} V_{\ul{i}}) &
A \Pi^i + B^{\ul{i}} \Phi^{i}_{\ \ul{i}} \\
- \alpha^{-1} W V_{\ul{i}} & \Phi^i_{\ \ul{i}}
\end{pmatrix}.
\end{align}
Note that the quantities~$\pi^{\ul{i}}$ and~$\Pi^i$ can be written in
terms of the lapse, shift and boost vectors
\begin{align} \label{eq:pieqn}
\pi^{\ul{i}} = W V^{\ul{i}} - W A^{-1} B^{\ul{i}}, &&
\Pi^{i} = W v^i - W \alpha^{-1} \beta^i.
\end{align}
Another important result we will need is that for a first order
evolution system in upper case coordinates of the form
\begin{align} \label{eq:uppercaseeqn}
\p_T \bm{u} = (A \bm{A}^{\ul{p}} + B^{\ul{p}} \bm{1})
\p_{\ul{p}} \bm{u} + A \bm{S},
\end{align}
where~$\bm{u}$ is the state vector, $\bm{A}^{\ul{p}}$ are the
principal matrices, and~$\bm{S}$ contains the source terms, can be
rewritten in terms of the lower case coordinates as
\begin{align} \label{eq:lowercaseeqn}
(\bm{1} + \bm{A}^{\ul{V}}) \p_t \bm{u} &= \alpha W^{-1} \left(
\bm{A}^{\ul{p}}(\varphi^{-1})^p_{\ \ul{p}} - (\bm{1} +
\bm{A}^{\ul{V}}) \Pi^p \right)  \p_p \bm{u} \nonumber \\
&\quad + \alpha W^{-1} \bm{S},
\end{align}
where~$\varphi^{\ul{i}}_{\ i} = {}^{\textrm{\tiny{(N)}}}
\gamma^{\ul{i}}_{\ \ul{\mu}} J^{\ul{\mu}}_{\ i}$ is called the
projected Jacobian and~$\bm{A}^{\ul{V}} \equiv\bm{A}^{\ul{i}}
V_{\ul{i}}$.

\subsection{DF in the generalized harmonic formulation}

In this subsection, we look at the generalized harmonic formalism
employed using the dual foliation approach. Our discussion will
closely follow~\cite{HilHarBug16} but with the addition of
the~$\gamma_1$ parameter, which because of the subtle asymptotics on
hyperboloidal slices was earlier hard-coded to vanish. We start
essentially with the first order GHG equations of~\cite{LinSchKid05}
which are in turn based on earlier work of Garfinkle~\cite{Gar01}.
With the~$\gamma_1$ parameter turned back on, they read
\begin{align}
  \p_T g_{\ul{\mu \nu}} &= (1 + \gamma_1) B^{\ul{i}} \p_{\ul{i}} g_{\ul{\mu \nu}}
  + A S^{(g)}_{\ul{\mu \nu}},\nonumber\\
  \p_T \Phi_{\ul{i \mu \nu}} &= B^{\ul{j}} \p_{\ul{j}} \Phi_{\ul{i \mu \nu}}
  - A \p_{i} \Pi_{\ul{\mu \nu}} + \gamma_2 A \p_{\ul{i}} g_{\ul{\mu \nu}}
  + A S^{(\Phi)}_{\ul{i \mu \nu}}, \nonumber \\
  \p_T \Pi_{\ul{\mu \nu}} &= \gamma_1 \gamma_2 B^{\ul{i}}
  \p_{\ul{i}} g_{\ul{\mu \nu}} + B^{\ul{i}} \p_{\ul{i}} \Pi_{\ul{\mu \nu}}
  - A \tinyN \gamma^{\ul{ij}} \p_{\ul{i}} \Phi_{\ul{j \mu \nu}}\nonumber\\
  &\quad+ A S^{(\Pi)}_{\ul{\mu \nu}}, 
\end{align}
where the source terms are given by
\begin{align}
  S^{(g)}_{\ul{\mu \nu}} &= - \Pi_{\ul{\mu \nu}}
  - \gamma_1 A^{-1} B^{\ul{i}} \Phi_{\ul{i \mu \nu}}, \nonumber \\
  S^{(\Phi)}_{\ul{i \mu \nu}} &= - \gamma_2 \Phi_{\ul{i \mu \nu}}
  + \frac{1}{2} N^{\ul{\alpha}} N^{\ul{\beta}} \Phi_{\ul{i \alpha \beta}}
  \Pi_{\ul{\mu \nu}} + \tinyN \gamma^{\ul{jk}} N^{\ul{\alpha}}
  \Phi_{\ul{i j \alpha}} \Phi_{\ul{k \mu \nu}}, \nonumber \\
  S^{(\Pi)}_{\ul{\mu \nu}} &= 2 g^{\ul{\alpha \beta}}
  \left(\tinyN \gamma^{\ul{ij}} \Phi_{\ul{i \alpha \mu}}
  \Phi_{\ul{j \beta \nu}} - \Pi_{\ul{\alpha \mu}} \Pi_{\ul{\beta \nu}}
  - g^{\ul{\delta \gamma}} \Gamma_{\ul{\mu \alpha \delta}}
  \Gamma_{\ul{\nu \beta \gamma}}\right) \nonumber \\
  &\quad - 2 \left(\nabla_{( \ul{\mu}} H_{{\ul{\nu}} )}
  + \gamma_3 \Gamma^{\ul{\alpha}}_{\ \ul{\mu \nu}} C_{\ul{\alpha}}
  - \frac{1}{2} \gamma_4 g_{\ul{\mu \nu}} \Gamma^{\ul{\alpha}}
  C_{\ul{\alpha}}\right) \nonumber \\
  &\quad - \frac{1}{2} N^{\ul{\alpha}} N^{\ul{\beta}} \Pi_{\ul{\alpha \beta}}
  \Pi_{\ul{\mu \nu}} - N^{\ul{\alpha}} \tinyN \gamma^{\ul{ij}}
  \Pi_{\ul{\alpha i}} \Phi_{\ul{j \mu \nu}} \nonumber \\
  &\quad + \gamma_0 \left[2 \delta^{\ul{\alpha}}_{\ ( \ul{\mu}}N_{\ul{\nu} )}
    - g_{\ul{\mu \nu}} N^{\ul{\alpha}}\right] C_{\alpha}
  - \gamma_1 \gamma_2 A^{-1} B^{\ul{i}} \Phi_{\ul{i \mu \nu}},
\end{align}
where we have
\begin{align}
  \Gamma_{\ul{\alpha \mu \nu}} &\equiv \tinyN \gamma^{\ul{i}}_{\ (\ul{\mu}|}
  \Phi_{\ul{i}|\ul{\nu} )\ul{\alpha}} - \frac{1}{2} \tinyN
  \gamma^{\ul{i}}_{\ \ul{\alpha}} \Phi_{\ul{i \mu \nu}}\nonumber\\
  &\quad+ N_{( \ul{\mu}} \Pi_{\ul{\nu})\ul{\alpha}}
  - \frac{1}{2} N_{\ul{\alpha}} \Pi_{\ul{\mu \nu}},
\end{align}
and the equations are subject to both the reduction constraints
\begin{align}
  C_{\ul{i\mu\nu}}=\p_{\ul{i}} g_{\ul{\mu\nu}}-\Phi_{\ul{i\mu\nu}},
  \label{eq:reduction_constraints}
\end{align}
and the GHG constraints
\begin{align}
  C_{\ul{\mu}} = g^{\ul{\alpha \beta}} \Gamma_{\ul{\mu \alpha \beta}}
  + H_{\ul{\mu}} = 0.
\end{align}
The functions~$H_{\ul{\mu}}$ are called the gauge source functions
which are functions of the coordinates and the metric. Now,
considering these evolution equations to be in the standard form of
Eqn.~\eqref{eq:uppercaseeqn}, we can easily obtain
the~$\bm{A}^{\ul{p}}$ matrices which is a slight modification from
that given in~\cite{HilHarBug16}
\begin{align}
\bm{A}^{\ul{p}} = \begin{pmatrix}
  \gamma_1 A^{-1} B^{\ul{p}} && 0 && 0 \\
  \gamma_2 \delta^{\ul{p}}_{\ \ul{i}} && 0 && -\delta^{\ul{p}}_{\ \ul{i}} \\
  \gamma_1 \gamma_2 A^{-1} B^{\ul{p}} && -\tinyN\gamma^{\ul{pj}} && 0
  \end{pmatrix}.
\end{align}
To avoid repeating the calculation in~\cite{HilHarBug16}, we observe
that by inclusion of the~$\gamma_1$ terms within the coordinate
change~\eqref{eq:lowercaseeqn} is straightforwardly done by modifying
the form given in~\cite{HilHarBug16} with the Sherman-Morrison
formula. Doing so we arrive at the lower case time evolution equations
\begin{align}
  \p_t g_{\ul{\mu \nu}} &= \left(\beta^p - \alpha v^p
  +\frac{ \gamma_1\alpha B^{\ul{p}}}{W(A+B^V)}
    (\varphi^{-1})^p{}_{\ul{p}}\right)
  \p_p g_{\ul{\mu \nu}}\nonumber \\
  &\quad + \alpha W^{-1} s^{(g)}_{\ul{\mu \nu}}, \nonumber \\
  d_t \Phi_{i\ul{\mu \nu}} &= \left(\beta^p \delta^{j}_{\ i}
  - \alpha v^p \delta^{j}_{\ i} + \alpha W^2 v_i (\mathbbmss{g}^{-1})^{pj}\right)
  d_{p} \Phi_{j\ul{\mu \nu}} \nonumber \\
  &\quad+\alpha W^{-1} \mathbbmss{g}^{p}_{\ i}
  \left(\gamma_2 \p_p g_{\ul{\mu \nu}} - \p_p \Pi_{\ul{\mu \nu}}\right)
  + \alpha W^{-1} s^{(\Phi)}_{i\ul{\mu \nu}}, \nonumber \\
  \p_t \Pi_{\ul{\mu \nu}} &= \beta^p \p_p \Pi_{\ul{\mu \nu}}
    - \alpha W
    (\mathbbmss{g}^{-1})^{p i} d_p \Phi_{i\ul{\mu \nu}}
    + \alpha W^{-1} s^{(\Pi)}_{\ul{\mu \nu}} \nonumber \\
  &\quad  -\gamma_2\left(\alpha v^p-\frac{ \gamma_1\alpha B^{\ul{p}}}{W(A+B^V)}
  (\varphi^{-1})^p{}_{\ul{p}}\right)
  \p_p g_{\ul{\mu \nu}}.
\end{align}
We use a shorthand notation which abbreviates the contraction with the
projected Jacobian
\begin{align}
d_{\mu} \Phi_{i \ul{\mu \nu}} = \varphi^{\ul{i}}_{\ i} \p_{\mu} \Phi_{\ul{i \mu \nu}},
\end{align}
and write~$B^V=B^{\ul{p}}V_{\ul{p}}$. The boost metric is
\begin{align}
\mathbbmss{g}_{ij} = \gamma_{ij} + W^2 v_{i} v_{j}.
\end{align}
In terms of the upper case sources, the lower case sources can be
written as
\begin{align}
  s^{(g)}_{\ul{\mu \nu}} &= \frac{S^{(g)}_{\ul{\mu \nu}}}{1+\gamma_1 A^{-1}B^V},
  \nonumber \\
  s^{(\Phi)}_{\ul{i \mu \nu}} &= S^{(\Phi)}_{\ul{i \mu \nu}}
  + W^2 V_{\ul{i}}\left(V^{\ul{j}} S^{(\Phi)}_{\ul{j \mu \nu}}
  - \gamma_2 S^{(g)}_{\ul{\mu \nu}} + S^{(\Pi)}_{\ul{\mu \nu}}\right), \nonumber \\
  s^{(\Pi)}_{\ul{\mu \nu}} &= \frac{\gamma_2S^{(g)}_{\ul{\mu \nu}}}{1+\gamma_1 A^{-1}B^V} 
  + W^2 \left(V^{\ul{j}}S^{(\Phi)}_{\ul{j \mu \nu}}
  -\gamma_2 S^{(g)}_{\ul{\mu \nu}} + S^{(\Pi)}_{\ul{\mu \nu}}\right).
\end{align}
We take~$\mathbbmss{s}_i$ to be an arbitrary spatial vector of unit
magnitude with respect to
it~$(\mathbbmss{g}^{-1})^{ij}\mathbbmss{s}_i\mathbbmss{s}_j=1$ and
define a projection operator orthogonal to~$\mathbbmss{s}_i$ by
\begin{align}
  \perpdsq^i{\!\!}_j=\gamma^i{}_j
  -(\mathbbmss{g}^{-1})^{ik}\mathbbmss{s}_k\mathbbmss{s}_j.
\end{align}
The characteristic variables of the system are given by
\begin{align}
u^{\hat{0}}_{\ul{\mu\nu}} &= g_{\ul{\mu\nu}}, \nonumber\\
u^{\hat{B}}_{i\ul{\mu\nu}} &= \perpdsq^j{\!\!}_i\Phi_{j\ul{\mu\nu}}
+ W\ \perpdsq^j{\!\!}_iv_j
(\Pi_{\ul{\mu\nu}}-\gamma_2 g_{\ul{\mu\nu}} ), \nonumber \\
u^{\hat{\pm}}_{\ul{\mu\nu}} &= \Pi_{\ul{\mu\nu}}
\mp \frac{W}{\sqrt{1+(v^{\mathbbmss{s}})^2}}
(\mathbbmss{g}^{-1})^{ij}\mathbbmss{s}_j
\Phi_{i\ul{\mu\nu}} - \gamma_2 g_{\ul{\mu\nu}},
\end{align}
with the corresponding characteristic speeds
\begin{align}
  &\beta^{\mathbbmss{s}}-\alpha\,v^{\mathbbmss{s}}
  +\gamma_1\frac{B^{\ul{p}}
  (\varphi^{-1})^p{}_{\ul{p}}\mathbbmss{s}_p}{A+\gamma_1B^V}\,,\nonumber\\
  &\beta^{\mathbbmss{s}}-\alpha\,v^{\mathbbmss{s}}\,,\nonumber\\
  &\beta^{\mathbbmss{s}}\pm
\alpha\sqrt{1+(v^{\mathbbmss{s}})^2}.
\end{align}

\subsection{DF scalar field}

For completeness, in this section, we compute the field equations for
the scalar field project employing the DF formalism. This calculation
is directly analogous to that in the last subsection. We start with
the first order form of the scalar field equations written in the
upper case coordinates
\begin{align}
\p_T \Phi &= B^{\ul{i}} \p_{\ul{i}} \Phi + A S^{(\Phi)}, \nonumber \\
\p_T \chi_{\ul{i}} &= B^{\ul{j}}\p_{\ul{j}} \chi_{\ul{i}} + A \p_{\ul{i}} \Pi
+ \gamma A \p_{\ul{i}} \Phi + A S^{(\chi)}_{\ul{i}}, \nonumber \\
\p_T \Pi &= B^{\ul{i}} \p_{\ul{i}} \Pi + A \tinyN \gamma^{\ul{ij}}
\p_{\ul{j}} \chi_{\ul{i}} + A S^{(\Pi)}.
\end{align}
where the source terms are given by
\begin{align}
S^{(\Phi)} &= \Pi, \nonumber \\
S^{(\chi)}_{\ul{i}} &= A^{-1} \chi_{\ul{j}} \p_{\ul{i}} B^{\ul{j}}
+ A^{-1} \Pi \p_{\ul{i}} A - \gamma \chi_{\ul{i}}, \nonumber \\
S^{(\Pi)} &= K \Pi + A^{-1} \chi_{\ul{i}} \tinyN\gamma^{\ul{ij}} \p_{\ul{j}} A
- \tinyN\gamma^{\ul{ij}} \tinythree \Gamma^{\ul{k}}_{\ \ul{ij}} \chi_{\ul{k}}.
\end{align}
The first order system is of the form given in
Eqn.~\eqref{eq:uppercaseeqn}, therefore we can construct the principal
matrices as
\begin{align}
\mathbf{A}^{\ul{p}} = \begin{pmatrix}
0 & 0 & 0 \\
\gamma \delta^{\ul{p}}_{\ \ul{i}} & 0 & \delta^{\ul{p}}_{\ \ul{i}} \\
0 & \tinyN\gamma^{\ul{pj}} & 0
\end{pmatrix}.
\end{align}
As mentioned in~\cite{HilHarBug16}, when the Lorentz factor~$W$ is
bounded, it is possible to invert the coefficient~$\left(\mathbf{1} +
\mathbf{A}^{\ul{V}}\right)$, which gives
\begin{align}
\left(\mathbf{1} + \mathbf{A}^{\ul{V}}\right)^{-1} = \begin{pmatrix}
1 & 0 & 0 \\
- \gamma W^2 V_{\ul{i}} & \tinyN \mathbbmss{g}^{\ul{j}}_{\ \ul{i}} & W^2 V_{\ul{i}} \\
- \gamma (W^2 - 1) & W^2 V^{\ul{j}} & W^2
\end{pmatrix},
\end{align}
where~$\tinyN \mathbbmss{g}^{\ul{j}}_{\ \ul{i}} = \tinyN
\gamma^{\ul{j}}_{\ \ul{i}} + W^2 V^{\ul{j}} V_{\ul{i}}$. Using this
information in Eqn.~\eqref{eq:lowercaseeqn}, we arrive at the
evolution equation for the scalar field variables in the lower case
coordinates
\begin{align}
  \p_t \Phi &= \left(\beta^p - \alpha v^p\right)\p_p \Phi
  + \alpha W^{-1} s^{(\Phi)}, \nonumber \\
  d_t \chi_{i} &= \left(\beta^p \delta^{j}_{\ i} - \alpha v^p \delta^{j}_{\ i}
  + \alpha W^2 v_i (\mathbbmss{g}^{-1})^{pj}\right) d_{p} \chi_j \nonumber \\
  &\quad + \alpha W^{-1} \mathbbmss{g}^{p}_{\ i} \left(\gamma \p_p \Phi
  + \p_p \Pi\right) + \alpha W^{-1} s^{(\chi)}_{i}, \nonumber \\
  \p_t \Pi &= \beta^p \p_p \Pi + \gamma \alpha v^{p} \p_p \Phi
  + \alpha W (\mathbbmss{g}^{-1})^{p i} d_p \chi_i \nonumber \\
  &\quad + \alpha W^{-1} s^{(\Pi)}.
\end{align}
Here again we use a shorthand notation which abbreviates contraction
with the projected Jacobian
\begin{align}
d_{\mu} \chi_{i} = \varphi^{\ul{i}}_{\ i} \p_{\mu} \chi_{\ul{i}}.
\end{align}
In terms of the upper case sources the lower case sources become
\begin{align}
s^{(\Phi)} &= S^{(\Phi)}, \nonumber \\
s^{(\chi)}_{\ul{i}} &= S^{(\chi)}_{\ul{i}} + W^2 V_{\ul{i}}\left(V^{\ul{j}}
S^{(\chi)}_{\ul{j}} - \gamma S^{(\Phi)} - S^{(\Pi)}\right), \nonumber \\
s^{(\Pi)} &= - \gamma S^{(\Phi)} - W^2
\left(V^{\ul{j}}S^{(\chi)}_{\ul{j}}-\gamma S^{(\Phi)} - S^{(\Pi)}\right).
\end{align}
The characteristic variables of the system are given by
\begin{align}
u^{\hat{0}} &= \Phi, \nonumber\\
u^{\hat{B}}_{j} &= \perpdsq^i{\!\!}_j \; \chi_{i} - W \ \perpdsq^i{\!\!}_j
\, v_{i}  \Pi, \nonumber \\
u^{\hat{\pm}} &= -\Pi \mp \frac{W}{\sqrt{1+(v^{\mathbbmss{s}})^2}}
(\mathbbmss{g}^{-1})^{ij}\mathbbmss{s}_j
\chi_{i} - \gamma \Phi,
\end{align}
with the corresponding characteristic speeds
\begin{align}
\beta^{\mathbbmss{s}}-\alpha\,v^{\mathbbmss{s}}\,,
\quad\beta^{\mathbbmss{s}}-\alpha\,v^{\mathbbmss{s}}\,,
\quad\beta^{\mathbbmss{s}}\pm
\alpha\sqrt{1+(v^{\mathbbmss{s}})^2}.
\end{align}
Here~$\mathbbmss{s}$ again denotes an arbitrary unit vector which is
normalized against the boost metric and is spatial with respect
to~$n^a$.

\section{DF Jacobians} \label{Section:DFJac}

In this paper, we are going to present the first numerical tests with
DF-GHG with an aim to not only change the spatial
coordinates~\cite{HemSchKid13} but also the foliation. DF-GHG is
implemented in 3d but for now we focus on spherical tests. To
demonstrate that everything in the code is correct, we implement a
list of Jacobians, some analytic and some that require the evolution
of additional fields. As a sanity check, the simplest Jacobian that we
implement is the identity Jacobian
\begin{align}
  t = T, && x^i = X^{\ul{i}},
\end{align}
which of course gives the correct result that we would expect in a run
without DF when the same gamma parameters are chosen for the job.
Although these Jacobians are primarily built for use in spherically
symmetric spacetimes, the implementation itself is made in our fully
3d code. This has the twin advantages that, using the Cartoon
method~\cite{AlcBraBru99,Pre04} for symmetry reduction, we can develop
and turn around simulations very quickly, but simultaneously end up
with code that can be used in a more general context. In the following
subsections we consider:
\begin{description}
\item[Analytic Jacobians] In these tests the two sets of coordinates
  are related by given closed form expressions. The new aspect is that
  in the past all simulations were performed under the simplifying
  assumption~$T=t$.
\item[Vanishing shift Jacobian] Close to the threshold of black hole
  formation in vacuum there are indications~\cite{HilWeyBru17} that
  popular choices of generalized harmonic coordinates form coordinate
  singularities. It is known that asymptotically flat spacetimes can
  always be foliated using a vanishing shift, which this Jacobian
  choice enforces.
\item[Areal radius Jacobian] In spherical symmetry there is a close
  relationship between the geometric radial coordinate and the null
  expansion. In this Jacobian we exploit this relationship to build
  coordinates in which (as long as we excise close enough to the
  apparent horizon) the excision boundary remains outflow for sure,
  and for which the apparent horizon is guaranteed to stay in the
  computational domain.
\item[DF-excision Jacobian] This Jacobian is an adjustment to the
  previous setup in which we use a solution to the eikonal equation to
  get tight control also over the incoming coordinate light speeds.
\end{description}
Alternative choices will be presented in future work.

\subsection{Analytic Jacobians}

First we consider the analytic Jacobian described by the following
relations
\begin{align}
  t = T, && x^i = f(t,r) X^{\ul{i}}.
\end{align}
Here we choose~$f_1(t,r)$ such that at~$t = 0$ and for large radius,
the upper case and the lower case coordinates match with each
other. We do not yet have provisions for the applying outer boundary
conditions in the DF case, so at large radius we require the
coordinates to change back to GHG where the usual GHG boundary
conditions in \verb|bamps| can be applied. A choice for $f$ which
satisfies these conditions is given by
\begin{align}
  f(t,r) = 1 + t^2 A_1 e^{-(r-r_0)^2} e^{-(t-t_0)^2},
\end{align}
where the Gaussian is centered such that its values approximately
reach machine precision or less near the outer boundary. Likewise, we
consider another Jacobian which is described by the following
relations
\begin{align} \label{eq:ajac2eqn}
	t = f(t,r) T, && x^i = X^{\ul{i}}.
\end{align}

\subsection{Vanishing shift Jacobian}

The vanishing shift Jacobian which keeps the lower case shift zero at
all times. Such a choice of coordinates may be useful when performing
simulations of gravitational collapse. First we choose
\begin{align}
t = T,
\end{align}
which makes some other quantities trivial, that is
\begin{align}
W=1, && \alpha = A, &&
V^{\ul{i}} = 0, && v^{i} = 0.
\end{align}
With these choices, we can write down the first of
Eqn.~\eqref{eq:pieqn} as
\begin{align}
\pi^{\ul{i}} = - A^{-1} B^{\ul{i}}.
\end{align}
This also simplifies the evolution equation for~$\phi^{\ul{i}}_{\ i}$
in Eqn.~\eqref{eq:jac} which can be obtained using Cartan's magic
formula~\cite{Hil15}
\begin{align} \label{eq:DiBibar}
\p_t \phi^{\ul{i}}_{\ i} = - D_i B^{\ul{i}} + \Lie_{\beta} \phi^{\ul{i}}_{\ i}.
\end{align}
The first term in the right hand side of the above equation can be
considered a source term, because by addition of the reduction
constraints, given in Eqn.~\eqref{eq:reduction_constraints}, all first
derivatives of metric components can be replaced by evolved variables,
whereas the second term should be ideally zero since we want the lower
case lapse to be zero. However, since we do not yet have outer
boundary conditions in the lower case coordinates, we will employ a
transition function approach. In this approach, we choose
\begin{align}
\beta^i = \Omega(r) B^{\ul{i}},
\end{align}
where~$\Omega$ is zero at small radii and transitions to one at large
radii. This allows us to apply the standard GHG boundary conditions
for the outer boundary.

The first source term in Eqn.~\eqref{eq:DiBibar} can be written as
\begin{align}
  \p_i B^{\ul{i}} = J^{\ul{k}}_{\ i} \p_{\ul{k}} B^{\ul{i}}. 
\end{align}
The upper case spatial derivatives of the upper case shift can then be
written down in terms of the lapse, shift, extrinsic curvature and the
Christoffel symbols, the expressions for which are given
below~\cite{Alc08}
\begin{align}
  \p_{\ul{m}} B^{\ul{l}} &= \Gamma^{\ul{l}}_{\ \ul{m0}} + B^{\ul{l}}
  \Gamma^{\ul{0}}_{\ \ul{m0}} + A K^{\ul{l}}_{\ \ul{m}} \nonumber \\
  &\quad- B^{\ul{n}}
  \Gamma^{\ul{l}}_{\ \ul{mn}} - B^{\ul{l}} B^{\ul{n}}
  \Gamma^{\ul{0}}_{\ \ul{mn}},
\end{align}
where further we use the expressions for the extrinsic curvature
\begin{align}
  K_{\ul{ij}} = - A \Gamma^{\ul{0}}_{\ \ul{ij}}.
\end{align}
The Lie derivative term in Eqn.~\eqref{eq:DiBibar} is only
non-vanishing for the subpatch where the transition happens and for
all outer subpatches. It can be written down as
\begin{align}
  \Lie_{\beta} \phi^{\ul{i}}_{\ i} =
  \Omega B^{\ul{k}} \p_k \phi^{\ul{i}}_{\ i} + \phi^{\ul{i}}_{j} \p_i \beta^j,
\end{align}
where
\begin{align}
  \p_i \beta^j = (\p_r \Omega) \Theta^{\ul{i}} B^{\ul{j}}
  + \Omega (\p_i B^{\ul{j}}).
\end{align}
where~$\Theta^i$ are functions of the angular coordinates which are
same for both the upper case and lower case coordinates and are
related to the Cartesian coordinates by the relation
\begin{align} \label{eq:thetadefn}
  x^j = r \Theta^j, && X^{\ul{j}} = R \Theta^{\ul{j}},
  && \Theta^i = \delta^i_{\ \ul{i}} \Theta^{\ul{i}}.
\end{align}
Putting all of this together, we can construct the required Jacobian
from which the inverse Jacobian can be computed numerically
\begin{align}
J =
\begin{pmatrix}
1 & -B^{\ul{i}} + \Omega B^{\ul{j}} \delta^j{\!}_{\ul{j}}\phi^{\ul{i}}_{\ j} \\
0 & \phi^{\ul{i}}_{\ i}
\end{pmatrix}.
\end{align}

\subsection{Areal radius Jacobian}

In this setup, we choose the Jacobian such that the lower case radial
coordinate is the areal radius. With this choice, we can show that the
position of the apparent horizon is located at the zero crossing of
the outgoing radial coordinate lightspeed. Now, since the position of
the apparent horizon in spherical symmetry in these coordinates can
only increase as the simulation progresses~\cite{Hay94,AshKri03}, if
the apparent horizon appears on the grid at the beginning of the
simulation, it must do so at later times. Consequently, as a result of
the weak cosmic censorship conjecture, the event horizon stays on the
numerical domain at all times. This ensures a successful `excision'
strategy.

Consider the upper case foliation whose spatial line element is given
by
\begin{align}
	ds^2 = L^2 dR^2 + \tinyN \gamma_{T} R^2  d\Omega^2.
\end{align}
Here~$L$ is called the length scalar, which is to the~$2+1$ split what
the lapse is to the~$3+1$ case. The relationship between the upper
case and the lower case radial coordinate is given by
\begin{align} \label{eq:reqetaR}
r = \eta (r, \tinyN \gamma_T) R,
\end{align}
where~$\eta$ is a function chosen such that for small $r$ the lower
case radial coordinate becomes the areal radius coordinate whereas for
large values of~$r$, it becomes the standard radial coordinate as
given by GHG. This ensures that normal GHG outer boundary conditions
can be applied for the system. A possible choice of~$\eta$ is of the
form
\begin{align}
\eta =  \sqrt{\tinyN \gamma_{T}} \chi(r) + 1(1-\chi(r)),
\end{align}
where~$\chi(r)$ is any suitable transition function varying from zero
to one with increasing~$r$. In principle, a hyperbolic tangent
function would serve the purpose but for reasons of rapid convergence,
we choose a low order polynomial function which transitions at the
penultimate subpatch. The functional form of $\chi(r)$ which
transitions from one to zero between~$r = r_0$ and $r = r_1$ can be
given by
\begin{align}
	\chi(r) = \begin{cases}
	1, &r < r_0,  \\
	1 - 3 a^2 (r-r_0)^2 - 2 a^3 (r-r_0)^3, &r_0 \leq r \leq r_1,  \\
	0, &r > r_1.
	\end{cases}
\end{align}
where $a = - 1 / (r_1 - r_0)$. The derivatives of $\eta$ are given by
\begin{align}
\tilde{\p}_r \eta = \sqrt{\tinyN \gamma_{T}} \chi'(r) - \chi'(r),
&&
\tilde{\p}_{(\tinyN\gamma_{T})} \eta = \frac{\chi(r)}{2 \sqrt{\tinyN \gamma_{T}}}.
\end{align}
Here the tilde on the partial derivatives means that the derivative
must be taken keeping the other argument constant. We shall now
construct the various components of the inverse Jacobian by noting
that in this case
\begin{align}
	\alpha = A, && V^{\ul{i}} = 0, && W = 1.
\end{align}
This information can be used to construct
the~$(J^{-1})^{0}_{\ \ul{i}}$ components of the inverse Jacobian. The
spatial components of the inverse Jacobian as given in
Eqn.~\eqref{eq:inversejac} following the relation given below
\begin{align}
  \Phi_{\ul{i}}^{\ j} = \p_{\ul{i}} x^j = \Theta^{\ul{j}} \p_{\ul{i}} r
  + r \p_{\ul{i}} \Theta^{\ul{j}},
\end{align}
using the relationship between~$r$ and~$\Theta^j$ as specified in
Eqn.~\eqref{eq:thetadefn}. The two terms of the above equation can be
evaluated using the fact that
\begin{align} \label{eq:pibarr}
  \p_{\ul{i}} r = \frac{\Theta^{\ul{i}} \eta
    + (r \tilde{\p}_{(\tinyN\gamma_{T})}\eta
    \p_{\ul{i}}\tinyN \gamma_{T})/\eta}{\kappa},
\end{align}
where
\begin{align}
  \kappa(r,\tinyN\gamma_T) \equiv 1 - \frac{r \tilde{\p}_r
    \eta}{\eta},
\end{align}
where we use Eqn.~\eqref{eq:reqetaR} and the fact that~$\p_{\ul{i}} R
= \Theta^{\ul{i}}$.  For the second term, we have
\begin{align}
  \p_{\ul{i}}X^{\ul{j}} = \delta^{\ \ul{j}}_{\ul{i}} =
  (\p_{\ul{i}} R) \Theta^{\ul{j}} + R \p_{\ul{i}} \Theta^{\ul{j}},
\end{align}
which gives
\begin{align}
  \p_{\ul{i}} \Theta^{\ul{j}} = \frac{\eta}{r}
  \left( \delta_{\ul{i}}^{\ \ul{j}} - \Theta^{\ul{i}} \Theta^{\ul{j}} \right).
\end{align}
We will now calculate the~$(J^{-1})^{i}_{\ \ul{0}}$ component which
can then be used to construct the time-space part of the inverse
Jacobian in Eqn.~\eqref{eq:inversejac}
\begin{align}
  (J^{-1})^{i}_{\ \ul{0}} = \p_T x^{i} = \Theta^{\ul{i}} \p_T r, &&
  \Pi^j = \frac{\Theta^{\ul{j}}\p_T r - B^{\ul{i}} \Phi^{j}_{\ \ul{i}}}{A}.
\end{align}
Now, the upper case time derivative of~$r$ can be computed from
Eqn.~\eqref{eq:reqetaR} in a straightforward manner
\begin{align} \label{eq:pTr}
  \p_T r = \frac{(r/\eta) \tilde{\p}_{(\tinyN\gamma_{T})} \eta
    \p_T \tinyN \gamma_{T}}{\kappa}.
\end{align}
For the sake of completeness, we also provide the upper case time and
spatial derivatives of~$\tinyN\gamma_{T}$ which are needed to
construct the above quantities. An expression for~$\tinyN\gamma_{T}$
can be written down in terms of the lapse, the determinant of the
Cartesian form of the metric and the length scalar as
\begin{align}
	\sqrt{-g_{\textrm{sph}}} = A L \sqrt{q},
\end{align}
where~$g_{\textrm{sph}}$ is the determinant of the metric in spherical
coordinates and~$q$ is the determinant of the two metric given by
\begin{align}
  \begin{pmatrix}
  R^2 \tinyN\gamma_{T} & 0 \\
  0 & R^2 \sin^2 \theta \tinyN\gamma_{T}
  \end{pmatrix}.
\end{align}
Using the fact that
\begin{align}
	\sqrt{-g_{\textrm{sph}}} = R^2 \sin \theta \sqrt{-g_{\textrm{cart}}},
\end{align}
we obtain an expression for~$\tinyN\gamma_T$ where the reference to
the Cartesian form of the metric is suppressed for the sake of
brevity
\begin{align}
\tinyN \gamma_T  = \frac{\sqrt{-g}}{A L}.
\end{align}
From here, the upper case time and spatial derivatives
of~$\tinyN\gamma_{T}$ can be obtained in a straightforward manner by
using the derivatives of~$\sqrt{-g}$, $A$ and~$L$. Using standard
results from the literature~\cite{Alc08}, we can compute time and
spatial derivatives of the square root of the determinant of the
metric
\begin{align}
  \p_T \sqrt{-g} = \sqrt{-g} \ \Gamma^{\ul{\mu}}_{\  \ul{0 \mu}}, &&
  \p_{\ul{i}} \sqrt{-g} = \sqrt{-g} \ \Gamma^{\ul{\mu}}_{\ \ul{i \mu}},
\end{align}
and also for the lapse
\begin{align}
  \p_T A = A (\Gamma^{\ul{0}}_{\ \ul{00}} - B^{\ul{m}} \Gamma^{\ul{0}}_{\ \ul{0m}}), &&
  \p_{\ul{i}} A = A (\Gamma^{\ul{0}}_{\ \ul{0i}} - B^{\ul{m}} \Gamma^{\ul{0}}_{\ \ul{i m}}).
\end{align}
The derivatives of the upper case length scalar~$L$ can be computed
from its definition
\begin{align}
L^{-2} = \tinyN\gamma^{\ul{ij}} \Theta^{\ul{i}} \Theta^{\ul{j}}.
\end{align}
To see that the apparent horizon is located at the zero crossing of
the outgoing radial coordinate lightspeed in area locking coordinates,
we consider the expression for the expansion which can be written
as~\cite{Alc08}
\begin{align}
H = \frac{1}{L}\left(\frac{2}{R}
+ \frac{1}{\tinyN\gamma_{T}}\p_R \tinyN \gamma_{T}\right)
- 2 \tinyN K^{\ul{\theta}}_{\ \ul{\theta}},
\end{align}
where~$\tinyN K_{\ul{ij}}$ is the extrinsic curvature in the upper
case foliation. A similar expression of course holds in an arbitrary
foliation. We have
\begin{align}
H \propto \left(\p_T + C_{+}^{R} \p_R\right) R^2 \tinyN\gamma_{T},
\end{align}
where~$C_{+}^{R}$ is called the outgoing radial coordinate lightspeed
and is defined as $C_{+} = - B^R + A/L$ where here and in the
following we suppress the label $R$. Now introducing area locking
coordinates~$(\mathring{T},\mathring{R} = R \sqrt{\tinyN\gamma_{T}})$,
we can write
\begin{align}
  H &\propto \left(\p_{\mathring{T}}
  + c_{+}^{\mathring{R}} \p_{\mathring{R}}\right) \mathring{R}^2, \nonumber \\
  &\propto 2 c_{+}^{\mathring{R}} \mathring{R}.
\end{align}
From the above expression, we see that in the case of the apparent
horizon, where the expansion is zero, $c_{+}^{\mathring{R}} = 0$
as~$\mathring{R}$ is greater than zero.

\subsection{Dual frame excision Jacobian}

As an addition to the previous strategy, which ensures the correct
sign of the outgoing radial coordinate lightspeed~$c_{+}$ at the inner
boundary of the simulation provided that we excise close enough to the
apparent horizon, we would like to exactly control the incoming radial
coordinate lightspeed~$c_{-}$, at least near the black
hole. If~$c_{-}$ can be set to~$-1$ exactly, this would avoid any
`artifical' coordinate redshift or blueshift as matter falls into the
event horizon. In this setup, the upper case
coordinates~$(T,X^{\ul{i}})$ are the generalized harmonic coordinates
whereas the lower case coordinates~$(t,x^i)$ are defined by the
Jacobian to be described shortly. The angular coordinates are kept to
be the same in both cases. The relationship between the upper case and
lower case radial coordinate is kept same as the previous
strategy. Furthermore, a new coordinate~$\mathring{v}$ is introduced
\begin{align} \label{eq:eikonaltransformation}
r = \eta (r, \tinyN \gamma_T) R, && \mathring{v} = \mathring{T} + r,
\end{align}
where~$\mathring{T} = \sqrt{\tinyN \gamma_{T}} \ T$. The only
condition that we impose on $\mathring{v}$ is that it be a null
coordinate, that is, it satisfies the eikonal equation
\begin{align} \label{eq:eikonaleq}
g^{ab} \nabla_a \mathring{v} \nabla_b \mathring{v} = 0,
\end{align}
The eikonal equation inside the transition region where~$\mathring{T}
= t$ can be expanded using the expression for the coordinate
lightspeeds along the radial direction
\begin{align} \label{eq:coordinatelightspeeds}
	c_{\pm} = - \beta^r \pm	\alpha/l ,
\end{align}
keeping in mind that~$l^{-2} = \gamma^{rr}$:
\begin{align}
\frac{1}{\alpha^2} (1 + c_{+}) (1 + c_{-}) = 0.
\end{align}
From the above expression, it can be clearly seen that when~$c_{+}$ is
not equal to~$-1$, $c_{-}$ takes the value of $-1$.

We have to ensure that near the outer boundary, the lower case time
coordinate reduces to the upper case time coordinate. This can be
achieved in a similar way as in the relationship between the radial
coordinates
\begin{align}
	t &= \eta(r, \tinyN \gamma_T) T, \nonumber \\
	  &= \mathring{T} \chi(r) + 1 (1 - \chi(r)) T,
\end{align}
Using Eqn.~\eqref{eq:eikonaltransformation}, we arrive at the final
relation between the upper case and the lower case time coordinate
\begin{align}
	t = \mathring{v} \chi - r \chi + (1 - \chi) T.
\end{align}
It is clear from the above expression that we have to evolve
derivatives of~$\mathring{v}$ and~$T$ to obtain the different
components of the inverse Jacobian. Instead of the components of the
Jacobian, we can choose to evolve an equivalent set of quantities
which are known as the `optical Jacobians'
\begin{align} \label{eq:ourvariables}
V^{-}_{\ul{i}} \equiv - \p_{\ul{i}} \mathring{v}, &&
E_{-} \equiv N^{\ul{\mu}} \p_{\ul{\mu}} \mathring{v}. 
\end{align}
In terms of these new variables, it is straightforward to show that
the eikonal equation in Eqn.~\eqref{eq:eikonaleq} can be rewritten as
\begin{align} \label{eq:eikonal2}
E^2_{-} &= \tinyN \gamma^{\ul{i j}} V^{-}_{\ul{i}} V^{-}_{\underline{j}}.
\end{align}
It is clear from the above expressions that the evolution equation
for~$E_{-}$ can be completely dropped in favor
of~$V^{-}_{\ul{i}}$. However, we choose to keep them since
$V^{-}_{\ul{j}}$ and~$E_{-}$ satisfy the eikonal equation, which can
be used to construct a constraint monitor.

\begin{figure*}[t] 
\centering
\includegraphics[width=\textwidth]{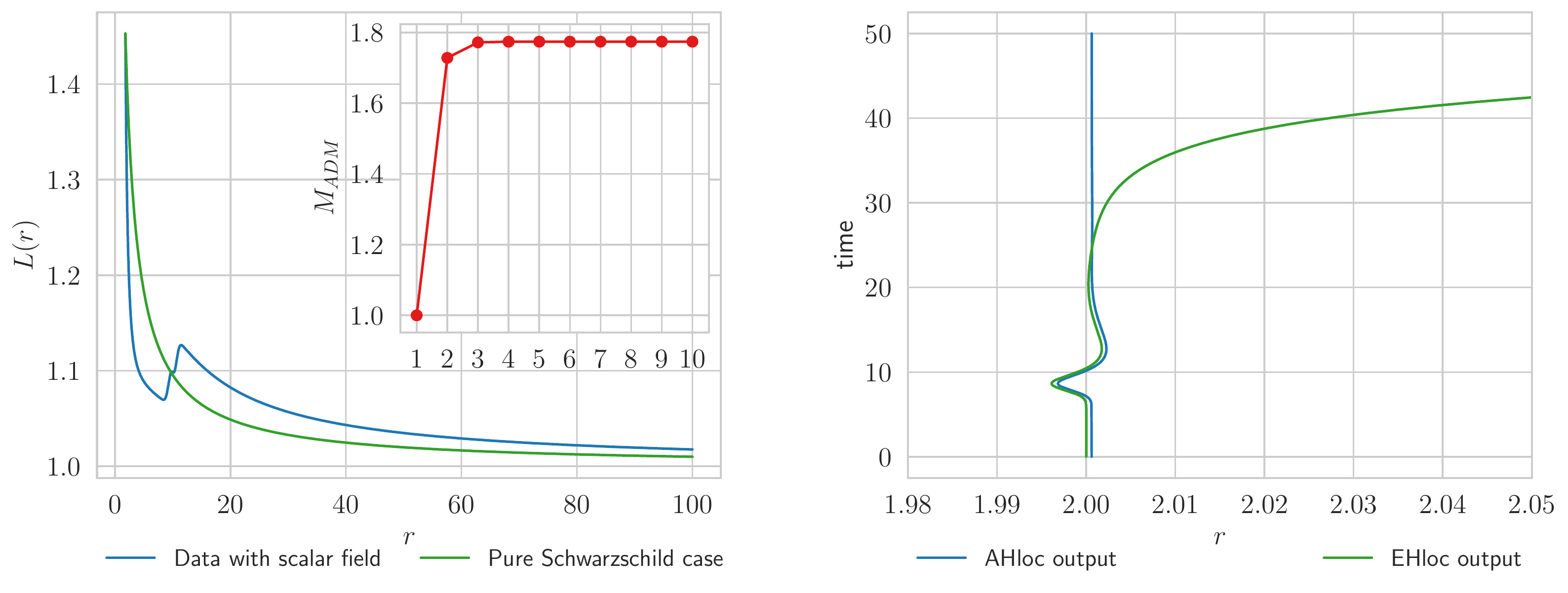}
\caption{Left: An example of constrained solved initial data with the
  blue line representing data corresponding to a non-zero scalar field
  and the green line representing the pure Schwarzschild case. The
  inset plot shows the values of $M_{\textrm{ADM}}$ which are
  generated during the iterative solve. Right: A comparison between
  the output of the event horizon locator and apparent horizon locator
  for a simulation with a lapse perturbation in the Schwarzschild
  spacetime. The deviation in the two outputs at later times
  demonstrates the event horizon locator trying to `find' the
  horizon. Here this effect is exaggerated because we chose a poor
  initial guess for the position of the event horizon on purpose.}
  \label{fig:EHAHgenerateIDplot}
\end{figure*}

We ask the reader to refer to~\cite{HilHarBug16} for a complete
derivation for the equations of motion for the optical Jacobians and
only provide a brief summary of the final equations here. The
evolution equations in the upper case coordinates can be written as a
set of advection equations such that this subsystem is minimally
coupled to the first order GHG system
\begin{align} \label{eq:uppercaseevolution}
\p_T V^{-}_{\ul{i}} &= (B^{\ul{j}} - A S^{\ul{j}}_{-})
\p_{\ul{j}} V^{-}_{\ul{i}} + A S^{(V^{-})}_{\ul{i}}, \nonumber \\
\p_T \ln E_{-} &= (B^{\ul{j}} - A S^{\ul{j}}_{-})
\p_{\ul{j}} \ln E_{-} + A S^{(E_-)},
\end{align}
where~$S^{\ul{j}}_{-} = E^{-1}_{-} V^{\ul{j}}_{-}$ and the source
terms are given by
\begin{align}
S^{(V^{-})}_{\ul{i}} &= A^{-1} V^{-}_{\ul{j}} \p_{\ul{i}} B^{\ul{j}}
+ S^{\ul{j}} \tinythree \Gamma^{\ul{k}}_{\ \ul{ij}} V^{-}_{\ul{k}} \nonumber \\
&\quad- A^{-1} E_{-} \p_{\ul{i}} A, \nonumber \\
S^{(E_-)} &= K_{S_- S_-}- \Lie_{S_-} \ln A.
\end{align}
Since our objective is to evolve the optical Jacobians in the lower
case `Cartesian' coordinates, we must transform the evolution
equations in Eqn.~\eqref{eq:uppercaseevolution} using
Eqns.~\eqref{eq:uppercaseeqn} and~\eqref{eq:lowercaseeqn}. To do this,
we compute~$(\varphi^{-1})^i_{\ \ul{i}}$ which can be written down in
terms of~$\Phi^i_{\ \ul{i}}$, $\Pi^i$ and $V_{\ul{i}}$,
\begin{align}
\left( \varphi^{-1}\right)^i_{\ \ul{i}} = \Phi^i_{\ \ul{i}} + \Pi^i V_{\ul{i}}.
\end{align} 
The complete principal matrix~$\mathbf{A}^{\ul{p}}$ associated with
the GHG variables and our new variables can be expressed as
\begin{align}
\mathbf{A}^{\ul{p}} = \begin{pmatrix}
\Lambda_1 & 0 \\
0 & \Lambda_2
\end{pmatrix},
\end{align}
where
\begin{align}
\Lambda_1 = \begin{pmatrix}
\gamma_1 A^{-1} B^{\ul{p}} && 0 && 0 \\
\gamma_2 \delta^{\ul{p}}_{\ \ul{i}} && 0 && -\delta^{\ul{p}}_{\ \ul{i}} \\
\gamma_1 \gamma_2 A^{-1} B^{\ul{p}} && -\tinyN\gamma^{\ul{pj}} && 0
\end{pmatrix},
\end{align}
and
\begin{align}
\Lambda_2 = \begin{pmatrix}
- S^{\ul{p}}_{-} & 0 \\
 0 & - S^{\ul{p}}_{-}
\end{pmatrix}.
\end{align}
Using the above expressions and Eqn.~\eqref{eq:lowercaseeqn}, the
evolution equations in the lower case coordinates can be written as
\begin{align}
\p_t V^{-}_{\ul{i}} &= (\beta^{j} - \alpha s^{j}_{-})
\p_{j} V^{-}_{\ul{i}} + \alpha W^{-1} s^{(V^{-})}_{\ul{i}}, \nonumber \\
\p_t \ln E_{-} &= (\beta^{j} - \alpha s^{j}_{-})
\p_{j} \ln E_{-} + \alpha W^{-1} s^{(E_-)},
\end{align}
where
\begin{align}
  s^i_{-} &= \frac{\left(\varphi^{-1}\right)^i_{\ \ul{i}}
    S^{\ul{i}}_{-}}{W \left(1 + W v_j
    \left(\varphi^{-1}\right)^j_{\ \ul{i}} S^{\ul{i}}_{-}\right)} + v^i,
\end{align}
and the lower case source terms are related to the upper case sources
as
\begin{align} 
  s^{(V^{-})}_{\ul{i}} &= \left(1 + W v_j
  \left(\varphi^{-1}\right)^j_{\ \ul{j}}
  S^{\ul{j}}_{-}\right)^{-1} S^{(V^{-})}_{\ul{i}}, \nonumber \\
  s^{(E_{-})} &= \left(1 + W v_j
  \left(\varphi^{-1}\right)^j_{\ \ul{j}}
  S^{\ul{j}}_{-}\right)^{-1} S^{(E_-)}.
\end{align}
It is straightforward to obtain the equations of motion for the other
two variables
\begin{align}
  \p_t R = J^{\ul{0}}_{\ 0}, && \p_t \mathring{v} =
  \frac{\alpha}{W} E_{-} - b^i W V_i
  - b^i \varphi^{\ul{j}}_{\ i} V^{-}_{\ul{j}},
\end{align}
where~$b^i = - \alpha v^i + \beta^i$. Finally, we can use the above
information to construct the different components of the optical
Jacobian. The~$\Phi_{\ \ul{i}}^{j}$ component are evaluated in the
same way as the previous case. Now
\begin{align}
  (J^{-1})^{\ 0}_{\ul{0}} &= \p_T t, \nonumber \\
  &= \left(A E_{-} - B^{\ul{j}} V^{-}_{\ul{j}}\right) \chi
  + \mathring{v} \p_r \chi \p_T r - \chi \p_T r \nonumber \\
  &\quad - r \p_r \chi \p_T r
  - T \p_r \chi \p_T r + (1-\chi), \nonumber \\
  (J^{-1})^{\ i}_{\ul{0}} &= \p_T x^i = \Theta^{\ul{i}} \p_T r, \nonumber \\
  (J^{-1})^{\ 0}_{\ul{i}} &= \p_{\ul{i}} t
  = - V_{\ul{i}}^{-} \chi + \mathring{v} \p_{\ul{i}} \chi
  - (\p_{\ul{i}} r) \chi \nonumber \\
  &\quad - r \p_{\ul{i}} \chi - T \p_{\ul{i}} \chi, 
\end{align}
where
\begin{align}
\p_{\ul{i}} \chi = (\p_r \chi) \Theta^{\ul{i}} \Phi^i_{\ \ul{i}}.
\end{align}
The~$\p_T r$ term in these equations can be simplified using
Eqn.~\eqref{eq:pTr}. To initialize the evolved quantities at the
beginning, we propose a choice which leads to the Jacobian being
identity initially. A choice of the evolved quantities is
\begin{align}
  V^{-}_{\ul{i}} = - \p_{\ul{i}} r, && \mathring{v} = r, && T = 0.
\end{align}
The choice of~$E_{-}$ is not independent but follows from the eikonal
equation.

Another important point to note is that we employ the cartoon method
to compute the~$y$ and~$z$ derivatives using Killing vectors. The
formula for doing this is provided below
\begin{align}
  \p_y V^{-}_{\ul{i}} &= h(x) \left(\delta^x_{\ \ul{i}} \delta^{\ul{j}}_{\ y}
  - \delta^y_{\ \ul{i}} \delta^{\ul{j}}_{\ x} \right) V^{-}_{\ul{j}}, \nonumber \\
  \p_z V^{-}_{\ul{i}} &= - h(x) \left(\delta^x_{\ \ul{i}} \delta^{\ul{j}}_{\ z}
  - \delta^z_{\ \ul{i}} \delta^{\ul{j}}_{\ x} \right) V^{-}_{\ul{j}}.
\end{align}
Note that~$h(x) = 1$ for the on-axis case and~$h(x) = 1/x$ otherwise.

Lastly, we briefly describe the constraint preserving outer boundary
conditions, the constraint being Eqn.~\eqref{eq:eikonal2}. At the
outer boundary, we choose~$\dot{V}^{-1}_{\ul{i}}$ to be equal to
zero. This requires a choice of~$\dot{E}_{-}$ which is given by
\begin{align}
  \dot{E}_{-} = - \frac{V_{\ul{i}}^{-} V_{\ul{j}}^{-}}{2 E_{-}}
  \tinyN \gamma^{\ul{i k}} \tinyN \gamma^{\ul{j l}}
  \ \p_T \tinyN \gamma_{\ul{kl}}. 
\end{align}

\section{Code setup}\label{Section:Code-Setup}

In this section we describe our numerical setup, initial data and
post-processing tools.

\subsection{Code overview}\label{section:bamps}

The~\verb|bamps| code~\cite{Bru11,HilWeyBru17,BugDieBer15,RueHilBug17}
is built for large scale, parallel numerical evolutions of hyperbolic
systems. Several different approximation schemes are implemented,
including DG schemes~\cite{BugDieBer15}, but here we use exclusively a
multidomain pseudospectral method to solve our first order symmetric
hyperbolic PDEs described in the previous sections. Each individual
numerical domain is called a subpatch. Within each subpatch spatial
derivatives are approximated using Chebyshev polynomials implemented,
as usual, by matrix multiplication. Data are communicated between
patches using a penalty method which is applied to the incoming
characteristic variables at each subpatch boundary. Our domain always
has a smooth timelike outer boundary at a fixed radial
coordinate~$r$. Because of this we need to apply boundary
conditions. These need to be constraint preserving, to control
undesirable gauge effects, and to control the physical behavior at the
boundary~\cite{SarTig12}. For now, to avoid introducing too many new
complications into the code at once, we choose Jacobians that
transition to the identity in a neighborhood of the outer
boundary. This allows us to recycle our boundary conditions for the
GHG formulation, essentially those of Rinne~\cite{Rin06a},
directly. For evolution in time we use a fourth order Runge-Kutta
method. Because we will be treating spherical spacetimes we use the
Cartoon method~\cite{AlcBraBru99,Pre04} to suppress two spatial
dimensions. With this reduction our tests are very fast, the longest
taking just a few minutes on a large desktop machine. We have tested
the implementation by evolving our spherical data with the full 3d
setup and obtain perfectly consistent results, and so do not discuss
these slower computations further. For a deeper technical description
of the code we direct the reader to~\cite{HilWeyBru17}.

\subsection{Initial data} \label{section:ID}

Our system involves a scalar field minimally coupled to the metric. To
evolve such a system, we must first solve for constraint preserving
initial data which can then be evolved using a combination of DF-GHG
and DF scalar field projects. We shall provide the necessary coupled
ordinary differential equations for the sake of completeness. Consider
the coordinates~$(r,\theta,\phi)$ in which the line element of the
spatial metric can be written as
\begin{align} \label{eq:spatialeqn}
ds^2 = l(r)^2 dr^2 + r^2 d\Omega^2,
\end{align}
where~$l$ is the lower case length scalar. The form of the extrinsic
curvature follows in a straightforward manner
\begin{align}
	K_{ij} =
	\begin{pmatrix}
	K_{rr}(r) & 0 & 0 \\
	0 & r^2 K_{T}(r) & 0 \\
	0 & 0 & r^2 \sin^2 \theta K_{T}(r)
	\end{pmatrix}.
\end{align}
We can now use this to obtain the Hamiltonian and the momentum
constraints which are given below respectively
\begin{align}
  &4 K_{T} K -6 K_{T}^2+\frac{2 \left(2 r l'+l^3-l\right)}{r^2 l^3}
  =8 \pi  \left(\frac{\Phi '^2}{l^2}+\Pi ^2\right), \nonumber \\
  &\frac{2 \left(r K_{T}'+3 K_{T}-K\right)}{r}=8 \pi  \Pi  \Phi ',
\end{align}
where~$K$ is the trace of the extrinsic curvature,~$\Phi$ is the
scalar field and~$\Pi$ is related to the time derivative of the scalar
field as
\begin{align}
  \Pi = - \frac{1}{\alpha} \left(\partial_t \Phi
  - \beta^i \partial_i \Phi \right).
\end{align}
From the Hamiltonian and momentum constraints we arrive at the ODEs
that we can solve using a Runge-Kutta method
\begin{align}
  \frac{dl}{dr} &= \frac{l }{2 r}\left(-2 r^2 K_{T} l^2 K +3 r^2 K_{T}^2 l^2
  +4 \pi  r^2 l^2 \Pi^2 \right. \nonumber \\
  &\quad\left. -l^2+4 \pi  r^2 \Phi '^2
  +1\right), \nonumber \\ 
  \frac{dK_{T}}{dr} &= \frac{-3 K_{T}+4 \pi  r \Pi \Phi '+ K}{r}.
\end{align}
This ODE is solved using an iterative method, with the trace of the
extrinsic curvature taken to be that in Schwarzschild with
the~$M_{\textrm{ADM}}$ mass taken to be one. The values of~$l(r)$
obtained in the first iteration is then used to construct the new ADM
mass, defined by
\begin{align}
  M_{\textrm{ADM}} = \frac{1}{2} r \left(l(r)^2-1\right),
  \quad r \rightarrow \infty
\end{align}
This is continued until the difference between the last and the second
last evaluation of the ADM mass meets a tolerance level. The spatial
metric quantities can then be reconstructed using
Eqn.~\eqref{eq:spatialeqn}, while the coordinate lightspeed $C_{+}$
constructed from
\begin{align}
	C_{+} = \frac{r - 2 M_{\textrm{ADM}}}{r + 2 M_{\textrm{ADM}}},
\end{align}
can be used to reconstruct the lapse and the shift
\begin{align}
  \alpha = \frac{l + C_{+} l}{2}, &&
  \beta_r = \frac{1 - C_{+}}{2}.
\end{align}
An example of the initial data solver in action is shown in the left
plot of Fig.~\ref{fig:EHAHgenerateIDplot}.
\subsection{Apparent and event horizon finders}

We require diagnostic tools for post-processing to ensure that the
excised region of spacetime remains inside the black hole event
horizon at all times during the numerical evolution. For this purpose
we use two tools, the apparent horizon, defined locally on a given
hypersurface and the event horizon which is a global property of the
spacetime.

The apparent horizon is defined as the outermost marginally outer
trapped surface on a given spatial hypersurface, that is, it is
defined by the vanishing of the expansion parameter of the outgoing
null geodesics. In spherical symmetry, the condition for the apparent
horizon is given by~\cite{Alc08}
\begin{align}
  H = \frac{1}{l}\left(\frac{2}{r}
  + \frac{1}{\gamma_{T}}\p_r \gamma_{T}\right) - 2 K^{\theta}_{\theta} = 0,
\end{align}
where the metric is represented in the lower case basis. This is
implemented in the \verb|AHloc| feature of \verb|bamps|. An
alternative and simpler way to find the apparent horizon in the area
locking coordinates is that the zero crossing of the lower case
outgoing coordinate lightspeed~$c_{+}$ corresponds to the position of
the apparent horizon.

We will now describe the implementation of a new event horizon finder
\verb|EHloc| for \verb|bamps|. The event horizon in general is a~$2+1$
null surface which is the boundary of the black hole region from which
no future pointing null geodesics can escape to null
infinity~$\mathcal{J}^{+}$ \cite{HawEll73}. Hence, one way to obtain
approximations of event horizons in numerical spacetimes is to
integrate null geodesics forward in time all over the numerical
domain. The disadvantage of this method is that simulations can only
be performed for a finite time and hence it is not straightforward to
find escaping null geodesics \cite{Tho06}. A more efficient algorithm
is to integrate outgoing null geodesics or null surfaces backwards in
time, since then the event horizon acts as an attractor of null
geodesics~\cite{Tho06,Die03}.

The geodesic method for integrating backwards in time is considered to
be the most accurate method and problems mentioned in the literature
like tangential drifting are not seen in
practice~\cite{CohPfeSch09}. Hence, this is the method we have used
for \verb|EHloc|.

Unlike \verb|AHloc|, it is essential for \verb|EHloc| to be run in
post-processing when the black hole is no longer ringing but is rather
Schwarzschild. In such cases, we can start from the last numerical
slice integrate the geodesic equation backwards
\begin{align}
  \frac{d^2 x^\alpha}{d\lambda^2}
  + \Gamma^{\alpha}_{\ \beta \gamma}
  \frac{dx^\beta}{d\lambda} \frac{dx^\gamma}{d\lambda} = 0,
\end{align}
where~$\lambda$ is the affine parameter and $x^\alpha$ is the
4-position of the geodesic. The initial conditions for the geodesic
are so chosen that it is outgoing. In spherical symmetry, the geodesic
equation can be represented by a set of coupled ordinary differential
equations~\cite{Boh16}
\begin{align}
  \frac{d\Pi_r}{dt} &= - \alpha_{,r} + (\alpha_{,r} \Pi^r
  - \alpha K_{rr} \Pi^r \Pi^r) \Pi_r + \beta^r_{\ ,r} \Pi_r \nonumber \\
  &\quad- \frac{1}{2} \alpha \gamma^{rr}_{\ ,r} \Pi_r \Pi_r, \nonumber \\
	\frac{dr}{dt} &= \alpha \Pi^r - \beta^r.
\end{align}
Here~$\alpha$, $\beta^i$ are the lapse and shift respectively,
$K_{ij}$ is the extrinsic curvature and~$\gamma^{ij}$ is the inverse
of the spatial metric. All quantities mentioned here are represented
in the lower case basis. The intermediate variable~$\Pi_r$ is related
to the momentum~$p_r = dx_{r}/d\lambda$ as
\begin{align}
\Pi_r \equiv \frac{p_r}{\sqrt{\gamma^{ij}p_i p_j}}.	
\end{align}
A Runge-Kutta integrator is used for performing the time stepping
while the data is loaded and then interpolated using Chebyshev
functions. A brief description of the grid and interpolation setup is
given below.

The data is written on every patch at the Gauss-Lobatto points
\begin{align} \label{eq:GL}
x_\alpha = - \cos \left(\frac{\pi \beta}{N - 1}\right),
\end{align}
where~$N$ is the number of points on each grid and~$\beta =
0,\ldots,N-1$. Chebyshev polynomials are used to perform the spectral
interpolation
\begin{align}
T_n (x) = \cos (n \cos^{-1} x).
\end{align}
These polynomials are defined in the interval~$[a,b]$ by a change of
variable:
\begin{align}
y \equiv \frac{x - \frac{1}{2}(b+a)}{\frac{1}{2}(b-a)}.
\end{align}
The coefficients of interpolation~$a_0,\ldots, a_{n-1}$ are found out
by solving:
\begin{align}
\begin{pmatrix} 
T_0(x_0) & \dots & T_{N-1}(x_0)\\
\vdots & \ddots & \\
T_0(x_{n-1}) &        & T_{N-1}(x_{N-1}) 
\end{pmatrix}
\begin{pmatrix}
a_0 \\
\vdots \\
a_{N-1}
\end{pmatrix} = 
\begin{pmatrix}
u_0 \\
\vdots \\
u_{N-1}
\end{pmatrix},
\end{align}
where~$u_0,\ldots,u_{N-1}$ are the given values at the~$N$ points. The
spatial derivatives are computed by a matrix
multiplication~\cite{HilWeyBru15}:
\begin{align}
(\partial_x u)_\alpha = \sum^{N-1}_{k = 0} D_{\alpha k} u_{k},
\end{align}
where~$D_{\alpha \beta}$ is the Gauss-Lobatto derivative matrix given
by
\begin{align}
  D_{\alpha \beta} = \begin{cases} \frac{-2(N-1)^2 + 1}{6},
  \ \alpha = \beta = 0, \\
  \frac{q_{\alpha}(-1)^{\alpha + \beta}}{q_\beta (x_\alpha - x_\beta)},
  \ \alpha \neq \beta, \\
  - \frac{x_\beta}{2(1-x^2_{\beta})}, \ \alpha = \beta = 1,\ldots,N-1, \\
  \frac{2(N-1)^2 + 1}{6}, \ \alpha = \beta = N-1, \end{cases}
\end{align}
where~$q_\alpha = 2$ at the boundary points and~$q_\alpha = 1$
elsewhere.

In practice, we do not compute the diagonal terms of the derivative
matrix but use the identity which gives the derivative matrix better
stability against rounding errors
\begin{align}
D_{\alpha \alpha} = - \sum^{N-1}_{k=0,k \neq \alpha} D_{\alpha k}.
\end{align}
The time interpolation on the data is performed using a linear
interpolation algorithm and a judicious choice of the number of points
needs to be taken into account. The number of data steps loaded into
memory also affects performance. However, both of these problems are
hardware specific and hence we do not go into details here. As sanity
checks, we test the event horizon finder with a multi-patch simulation
of the Schwarzschild spacetime and another with a gauge
perturbation. These results have also been compared with the output of
the apparent horizon finder and seen to be in good agreement. The
gauge perturbation case with both \verb|EHloc| and \verb|AHloc|
outputs are shown in the right plot of
Fig.~\ref{fig:EHAHgenerateIDplot}.

\section{Numerical results}\label{Section:Numerics}

\begin{figure*}[t] 
\centering \includegraphics[width=\textwidth]{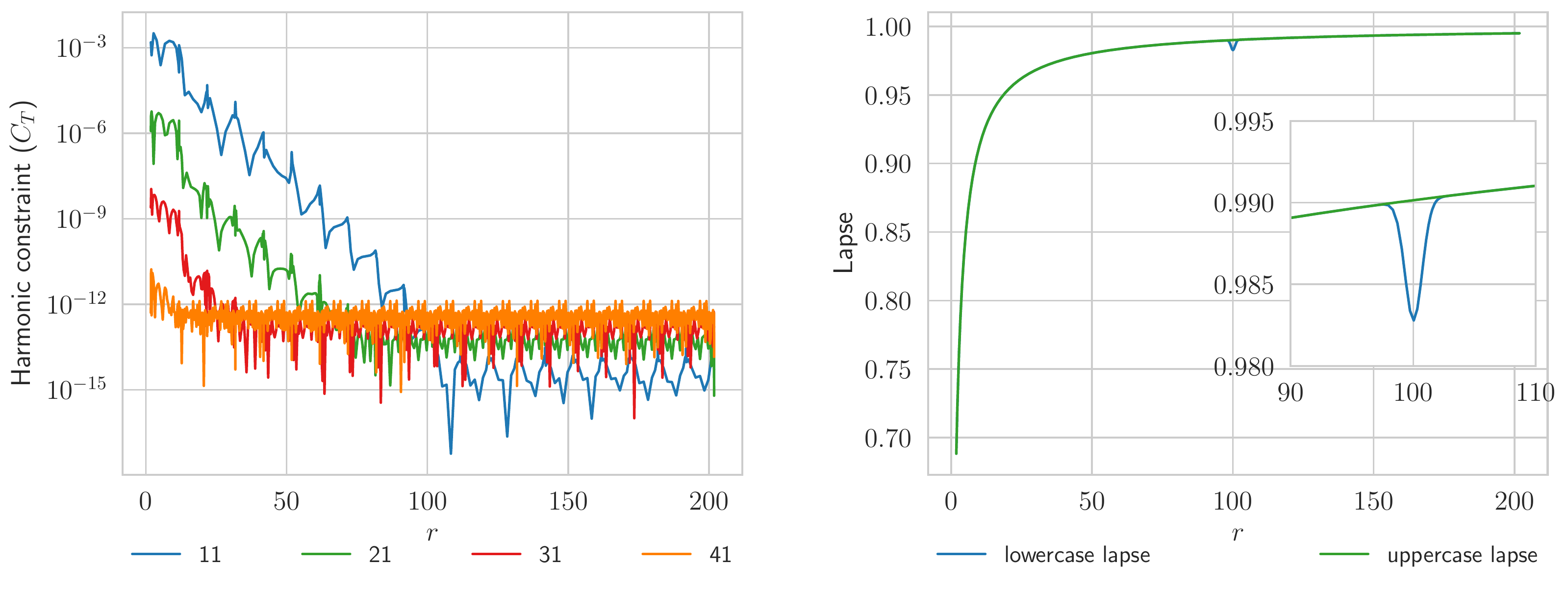}
\caption{Left: A convergence test performed with the time component of
  the harmonic constraints when the analytic Jacobian given in
  Eqn.~\eqref{eq:ajac2eqn} is considered. The numbers in the legend
  correspond to the number of points per patch considered. Right: A
  comparison between the lower case and the upper case lapse at time
  $t \simeq 5$ for the same set of
  simulations.}\label{fig:analyticJac}
\end{figure*}

\subsection{Tests with analytic Jacobians}

We begin our numerical experiments by first testing out our
implementation of the analytic Jacobians. As initial data, we choose
the metric components to be those of the Schwarzschild spacetime in
Kerr-Schild coordinates. We perform simulations for both the analytic
Jacobians keeping the function $f$ to be
\begin{align}
  f = 1 + 0.0001 t^2 e^{-(r-100)^2} e^{-(t-5)^2}.
\end{align}
The numerical domain for these simulations are from~$r \in
[1.8,201.8]$ and they are performed at $4$ different resolutions
starting from $20$ patches, $11$ points and increasing the number of
points by $10$ in each case. We finally plot the harmonic constraints
\begin{align}
  C_{\ul{\alpha}} = H_{\ul{\alpha}}
  + g^{\ul{\beta \gamma}} \ \Gamma_{\ul{\alpha \beta \gamma}},
\end{align}
with radius at four different resolutions and find that the
constraints converge with increasing resolution. A plot of such a
convergence test is provided in Fig.~\ref{fig:analyticJac}. We also
perform tests of the time derivatives of the harmonic constraints
given by
\begin{align}
  F_{\ul{\alpha}} \simeq \p_N H_{\ul{\alpha}} + g^{\ul{\beta \gamma}}
   \p_N \Gamma_{\ul{\alpha \beta \gamma}}
  - \Gamma_{\ul{\alpha}}^{\ \ul{\beta \gamma}} \p_N g_{\ul{\beta \gamma}},
\end{align}
where~$\simeq$ denotes equality up to the combinations of the
reduction constraints and~$\p_N\equiv N^{\ul{\alpha}}\p_{\ul{\alpha}}$
(see~\cite{LinSchKid05,HilWeyBru15} for details). We find that they
also converge with increasing resolution.

\subsection{Tests with the vanishing shift Jacobian}

\begin{figure*}[t] 
\centering
\includegraphics[width=\textwidth]{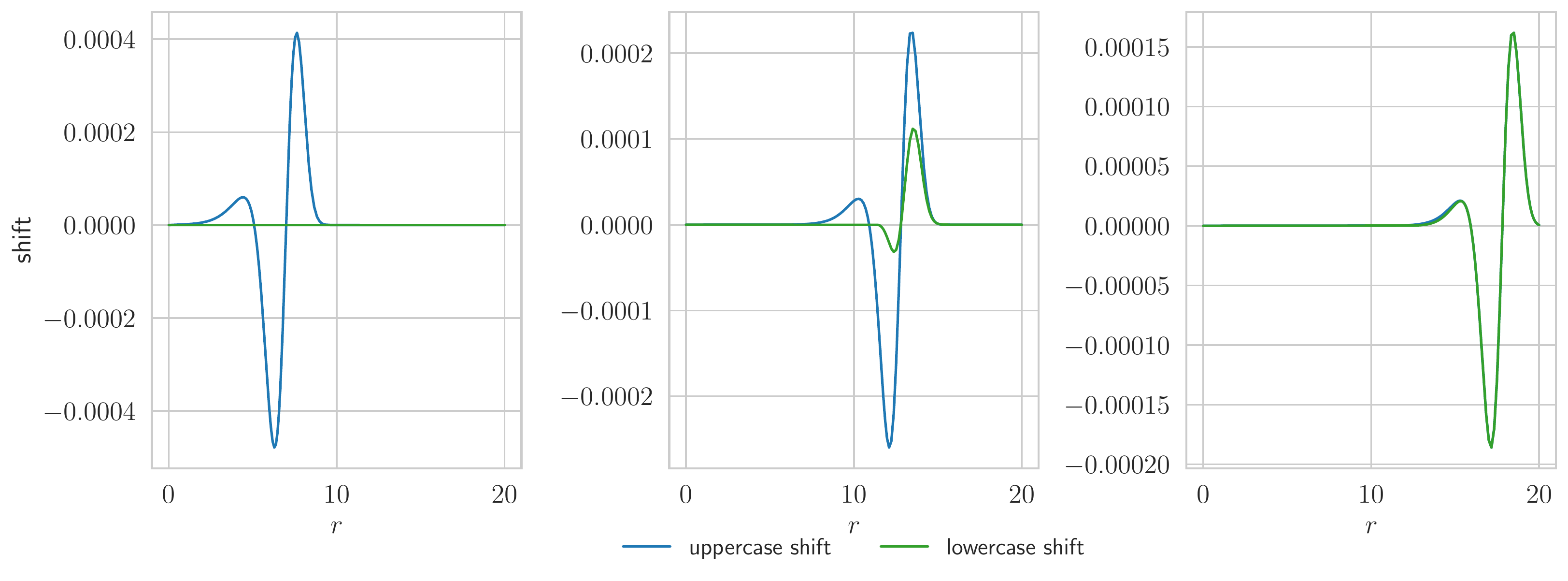}
\caption{A snapshot of the upper case shift and the lower case shift
  at three different times given by $t \simeq 7.7$, $13.5$ and $18.5$
  (in units of $M$). In the first figure, we see as expected
  irrespective of the upper case shift, the lower case shift is
  vanishingly small inside the transition region. In the second
  figure, we demonstrate the effect of the transition region on the
  lower case shift. In the third figure, we see the form of the lower
  case shift mostly outside the transition region where it is expected
  to agree with its upper case counterpart.}\label{fig:killshiftplot}
\end{figure*}

Another numerical experiment we perform is to keep the lower case
shift to be zero by a suitable choice of Jacobian. This experiment is
performed on the Minkowski spacetime by adding a Gaussian gauge wave
of the form
\begin{align}
\alpha &= 1 + G' e^{-w'(r-r_0)^2},
\end{align}
where the parameters of the perturbation are given by
\begin{align}
G' = 0.2, && r_0 = 0, && w' = 1.
\end{align}
As has been demonstrated in the calculations of
section~\ref{Section:DFJac}, the upper case and the lower case
coordinates match in the outermost subpatch of the simulation while
the penultimate subpatch serves as the transition region. This can be
seen clearly in the plots in Fig.~\ref{fig:killshiftplot} where the
upper case shift is represented by the blue curve and the lower case
shift is represented by the green curve. The lower case shift is
successfully kept to zero in patches inside the transition zone, while
outside the transition zone, it is seen to agree with the upper case
shift. A convergence test is also performed considering the reduction
constraints, the harmonic constraints and the time derivatives of the
harmonic constraints and we see convergence with increase in
resolution, as is expected.

\subsection{Tests with the areal radius Jacobian}

\begin{figure*}[t] 
\centering
\includegraphics[width=\textwidth]{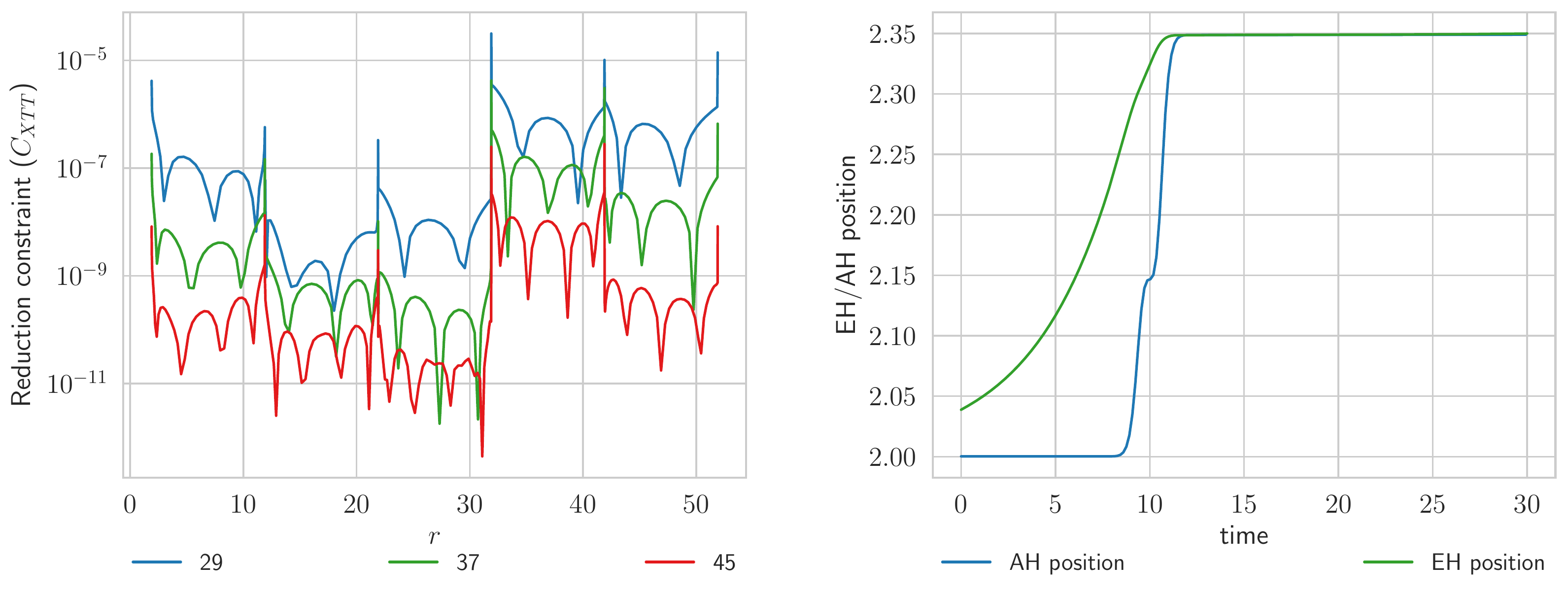}
\includegraphics[width=\textwidth]{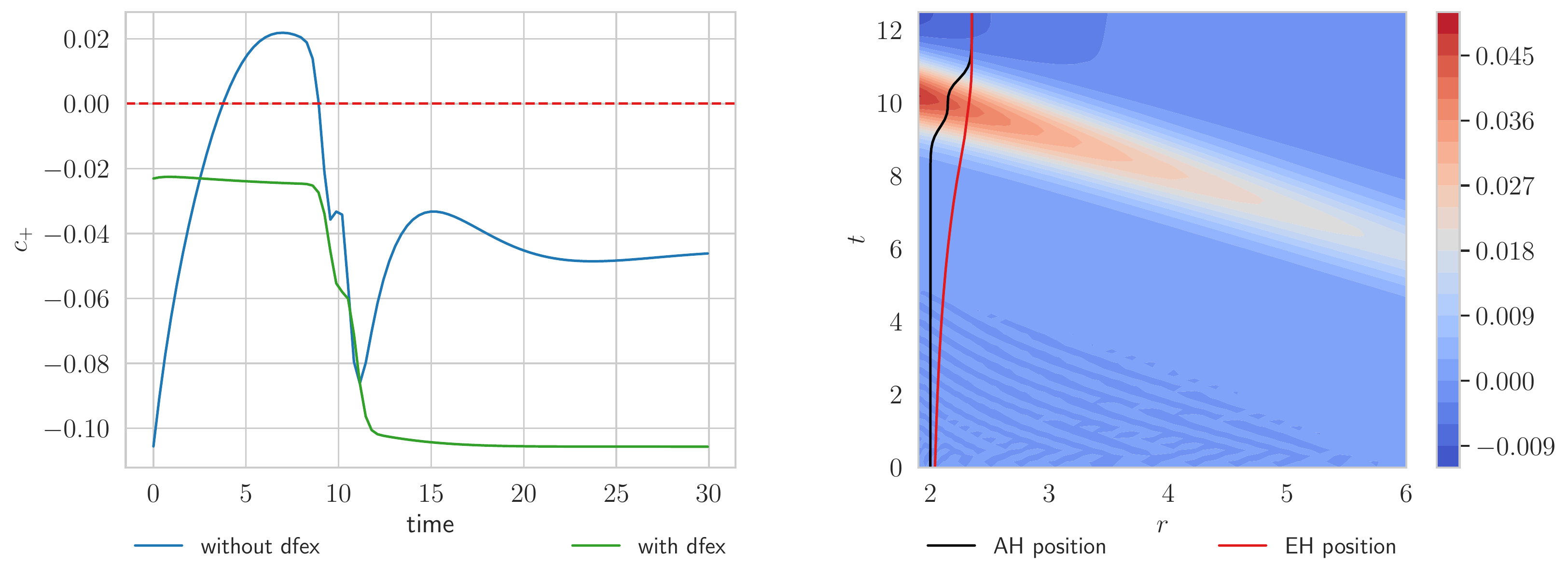}
\caption{Top row, left: A convergence test performed with the~$XTT$
  component of the reduction constraints at~$t = 30 M$ performed at
  three different resolutions, the number of points per patch being
  mentioned in the legend. These simulations are performed with the
  GHG, DF and scalar field projects. Top row, right: The position of
  the apparent horizon and the event horizon as a function of time in
  one of these simulations with DF, GHG and scalar field. The position
  of the apparent horizon, also where~$c_{+}$ crosses zero is seen to
  increase monotonically as a function of time. Bottom row, left: A
  plot showing the value of the outgoing radial coordinate
  lightspeed~$c_{+}$ at the inner boundary of two simulations, one
  performed with DF-excision and another performed without it. The
  simulation performed with DF-excision switched on has a negative
  value of~$c_{+}$ at the inner boundary throughout the simulation
  while the non-DF simulation fails in maintaining that. Bottom row,
  right: The time evolution of the scalar field along with the
  position of the event horizon and the apparent horizon.}
\label{fig:arealradiusplot}
\end{figure*}

We now perform simulations of a massless scalar field minimally
coupled to general relativity in spherical symmetry. As a first try,
we evolve the spacetime in generalized harmonic
coordinates~\cite{LinSchKid05} using the old excision setup, that is,
there are no boundary conditions placed at the inner boundary which is
expected to be outflow at the beginning of the simulation. We also
monitor the signs of the two radial coordinate lightspeeds as given in
Eqn.~\eqref{eq:coordinatelightspeeds} at the inner boundary of the
simulation at all times. We set the scalar field data initially to be
of the following profile
\begin{align} \label{eq:scalarID}
	\Phi &= \frac{C}{r} e^{-(r-r_0)^2/\sigma^2}, \nonumber \\
	\Pi &= - \frac{2 C}{r \sigma^2} (r - r_0) e^{-(r-r_0)^2/\sigma^2}.
\end{align}
The specific parameters which are chosen for the run are
\begin{align}
	C = 0.1, && r_0 = 11.9, && \sigma = 1.
\end{align}
Using these parameters, we then solve for the spatial metric,
extrinsic curvature, lapse and shift in the initial data using the
method prescribed in section~\ref{section:ID}.

A plot of the outgoing radial coordinate lightspeed~$C_{+}$, as can be
seen in the bottom row, left of Fig.~\ref{fig:arealradiusplot} shows
that it assumes a positive sign at the inner boundary for some time
during the simulation. This indicates that the excision strategy has
failed as the inner boundary has not remained an outflow boundary
during those times. This experiment clearly demonstrates that for
certain configurations of the matter content, the existing excision
strategy is unsuccessful.

We now perform the same experiment, but this time we switch on DF and
the areal radius Jacobian as described in
section~\ref{Section:DFJac}. As described before, the use of the areal
radius ensures that the position of the apparent horizon can only
monotonically increase with time, which ensures that if it is
initially located inside the numerical domain, it shall do so at all
times. Like before, the lightspeeds at the inner boundary are
monitored at all times. It can be clearly seen, from the green line in
the bottom left of Fig.~\ref{fig:arealradiusplot} that the value
of~$C_{+}$ at the inner boundary remains negative at all times during
the simulation thereby indicating that the new excision strategy is
successful. Although not seen in the plot, the~$C_{-}$ lightspeed,
while its value fluctuates, remains negative at all times for both the
DF and the non-DF case. This fact can be seen from
Eqn.~\eqref{eq:coordinatelightspeeds} which shows that if~$C_{+}$
remains negative at all times, so must the value of~$C_{-}$. We also
perform simulations with different quantities of scalar field content
and find many other cases where the new method proves to be successful
where the old one does not.

With the parameters which have been provided above, we perform
simulations at three different resolutions, having~$29$, $37$ and~$45$
points per patch and plot the reduction constraints which are defined
by~\eqref{eq:reduction_constraints} and can be computed in the lower
case coordinates from
\begin{align}
  C_{\ul{i \alpha \beta}} = \left( \varphi^{-1}\right)^k_{\ \ul{i}}
  \p_k g_{\ul{\alpha \beta}} + V_{\ul{i}} \Pi_{\ul{\alpha \beta}}
  - \Phi_{\ul{i \alpha \beta}},
\end{align}
as a function of space for a given value of time. We see convergence
with increase in resolution, as can be seen in the top left plot of
Fig.~\ref{fig:arealradiusplot}.

We also employ our event horizon and apparent horizon finders to track
the location of the horizons. A superposition of the output of the two
finders is provided at the top right of
Fig.~\ref{fig:arealradiusplot}. As expected the apparent horizon grows
monotonically with time from~$2 M$ to~$\simeq 2.35 M$. The event
horizon also shows a monotonic behavior in these coordinates.

\begin{figure*}[t] 
\centering
\includegraphics[width=\textwidth]{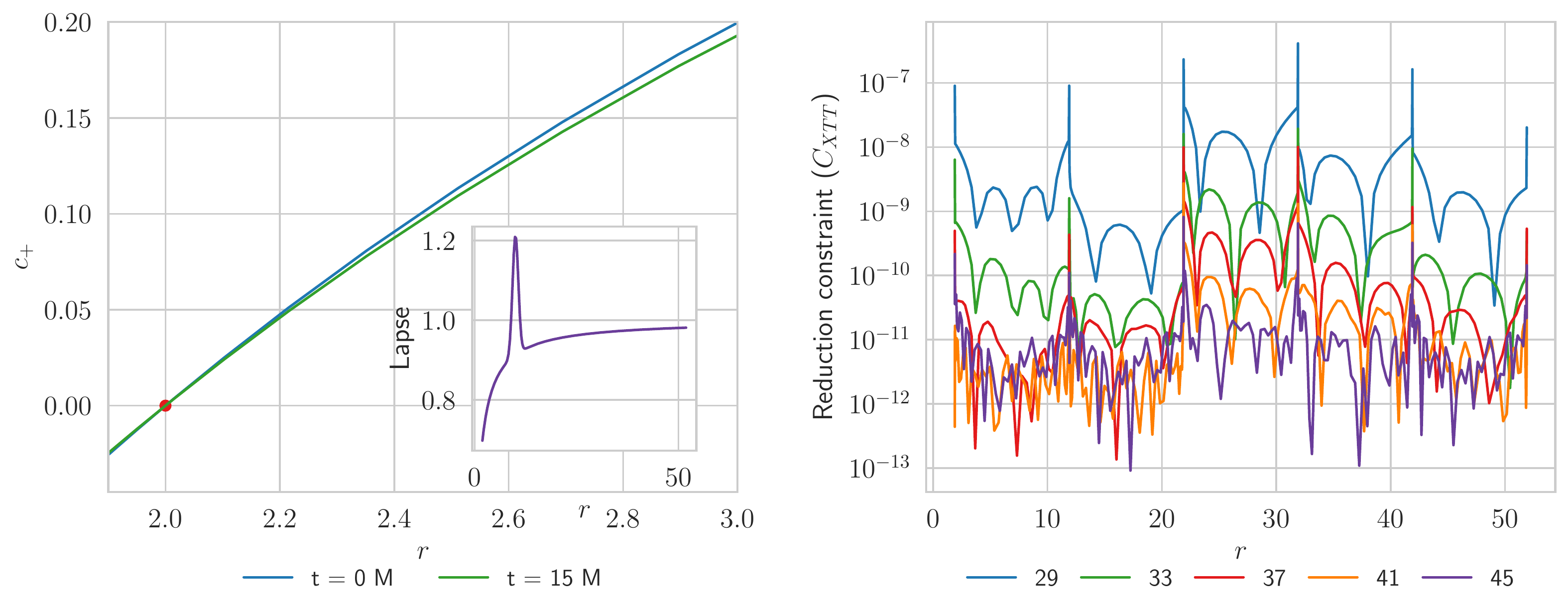}
\caption{Left: A demonstration of the fact that the position of the
  apparent horizon does not change when hit by a lapse
  perturbation. In the figure, the outgoing radial coordinate
  lightspeed~$c_{+}$ is shown at two different times and it is seen
  that the zero crossing remains at~$2 M$ throughout the
  simulation. The inset plot shows the lapse perturbation in the
  initial data. Right: A convergence plot of the~$XTT$ component of
  the reduction constraints performed with five different resolutions
  of the same lapse perturbation simulation. The legend shows the
  number of points per patch.}\label{fig:lapsepertplot}
\end{figure*}

Another experiment we perform involves evolving the Schwarzschild
spacetime with a lapse perturbation
\begin{align}
	\alpha &= \alpha' + H' e^{-w'(r-r_0)^2}, \nonumber \\
	\p_{i} \alpha &= \p_{i} \alpha'
        - 2 H' w' (r - r_0) e^{-w'(r-r_0)^2} \frac{x_i}{r},
\end{align}
where~$\alpha'$ is the natural lapse associated with the Schwarzschild
metric in Kerr-Schild coordinates. The specific parameters which are
chosen for this experiment are
\begin{align}
	H' = 1, && r_0 = 10, && w' = 1.
\end{align}
The lapse perturbation is shown in the inset plot of
Fig.~\ref{fig:lapsepertplot}. We observe the zero crossing of the
outgoing radial coordinate lightspeed~$C_{+}$ as this corresponds to
the position of the apparent horizon. At the beginning of the
simulation, this stays at~$2 M$ and continues to remain so throughout
the entire duration of the simulation. This is shown by the blue and
green lines in the left plot of Fig.~\ref{fig:lapsepertplot} which
correspond to the lightspeed at the beginning and at the end of the
simulation. This is indeed the desired behavior since the position of
the apparent horizon should not change as there is no physical
perturbation.

\subsection{Tests with the DF-excision Jacobian}

\begin{figure*}[t] 
\centering
\includegraphics[width=\textwidth]{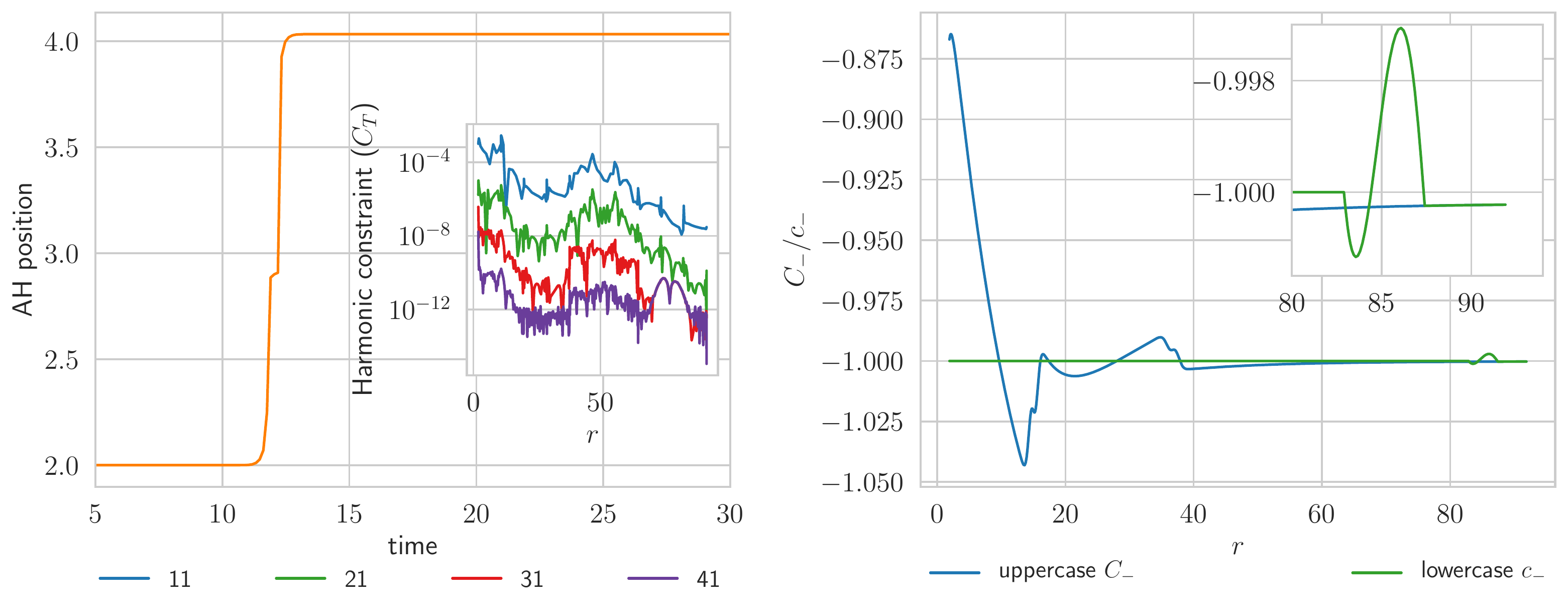}
\caption{Left: A demonstration showing the growth of the apparent
  horizon by~$\simeq 101 \%$ by accreting scalar field into the black
  hole. The inset plot shows a convergence test with the time
  component of the harmonic constraint. The legend shows the number of
  points per patch. Right: A plot of the incoming radial coordinate
  lightspeed~$C_{-}/c_{-}$ in both the upper case and the lower case
  coordinates with the lower case result shown to be~$-1$ inside the
  transition region. The inset plot zooms in on the transition region
  and shows that the upper case and lower case speeds agree at the
  outermost subpatch.}\label{fig:dfexeikonal}
\end{figure*}

Finally, we perform numerical experiments with the dual foliation
eikonal Jacobian. The numerical setup again consists of a massless
scalar field minimally coupled to general relativity. Control of
the~$C_{+}$ radial coordinate lightspeed is borrowed from the
treatment in the previous section. In this section, our goal is also
to control the ingoing radial coordinate lightspeed~$C_{-}$ to be
identically~$-1$ inside the transition region. This would prevent any
redshift or blueshift of the scalar field pulse as it falls into the
event horizon. We prepare initial data using our initial data solver
for scalar field of the type given by Eqn.~\eqref{eq:scalarID} with
the parameters given by
\begin{align}
C = 0.21, && r_0 = 15, && \sigma = 1.
\end{align}
Our principal objective in this experiment is to grow the apparent
horizon as much as possible by letting accreting scalar field fall
into the black hole horizon. For this specific choice of parameters,
we see that the apparent horizon position grows from~$\simeq 2 M$
to~$4.03 M$ thereby registering $\simeq 101 \%$ increase. This
increase, which is monotonic with time is demonstrated clearly in the
left plot of Fig.~\ref{fig:dfexeikonal}. We also perform convergence
tests by considering simulations with~$11$, $21$, $31$ and~$41$ points
per patch and with each simulation containing~$10$ patches. Plots of
the reduction constraints, harmonic constraints and the time
derivatives of the harmonic constraints all demonstrate convergence
with increasing resolution as is expected. As a demonstration, the
inset plot of the right hand side of Fig.~\ref{fig:dfexeikonal} shows
the convergence of the harmonic constraints.

Finally, we look at the outgoing radial coordinate lightspeed in both
the upper case and the lower case coordinates, as can be seen from the
right plot of Fig.~\ref{fig:dfexeikonal}. The upper case~$C_{-}$ is
seen to vary freely while our method ensures that the lower
case~$c_{-}$ is strictly kept to be equal to~$-1$ throughout the
entire simulation inside the transition region. As can be seen from
the inset plot of the same figure, the two lightspeeds disagree in the
transition patch but they do agree in the outermost patch as expected.

\section{Conclusions}\label{Section:Conclusions}

In this paper, we have presented the first implementation of the
DF-GHG formulation, together with the DF-scalar field. The
implementation was made within the \verb|bamps| code. We performed a
battery of tests involving several Jacobians, but with an emphasis on
black hole excision. Although the tests performed are in spherical
symmetry as proof of concept and also for reasons of efficiency, the
whole implementation itself was made in the full $3+1$ setting. In
addition to this, we introduced our event horizon finding code
\verb|EHloc|.

To test the newly written DF-GHG project, we have performed elementary
tests with two analytic Jacobians, in one of which we consider two
different foliations for the upper case and lower case
coordinates. After this, we tested the vanishing shift Jacobian,
another important case in which the lower case shift is kept zero at
all times, which we expect to be helpful while considering cases of
gravitational collapse and black hole formation.

Finally, we considered the two most important Jacobians for our black
hole excision work, the areal radius Jacobian and the DF-excision
Jacobian. In the areal locking case, we saw that when the lower case
coordinates are made to include the areal radius, the apparent horizon
is located at the zero-crossing of the outgoing radial coordinate
lightspeed~$c_{+}$. By basic results for dynamical horizons, these
coordinates also have the special property that the position of the
apparent horizon cannot decrease. Thus if the apparent horizon is
initially located on the numerical grid, it stays so throughout the
simulation. Assuming the weak cosmic censorship conjecture, we then
ensure that the event horizon, being located outside the apparent
horizon, also remains on the numerical grid. In the case of the
DF-excision Jacobian, we carefully control the ingoing radial
coordinate lightspeed~$c_{-}$ to be equal to~$-1$ identically while
enforcing the previous condition for the outgoing
speed. Controlling~$c_{-}$ to be a constant everywhere on the
numerical grid, barring the transition region and outside, ensures
that the redshift or blueshift that potentially arises out of using
`artificial' coordinates is avoided. We performed a series of tests on
the Minkowski spacetime with lapse perturbations or perturbed
Schwarzschild spacetimes by employing a combination of the DF, DF-GHG
and DF-scalarfield projects. These tests were validated by performing
several convergence tests, which demonstrate clean spectral
convergence.

The excision setup presented here is of course {\it highly}
specialized when compared with the full control system approach used
in the SpEC code. That said it provides precisely the functionality
needed for our near-term work, and has the advantage that we use only
pointwise, rather than quasilocal manipulation of our variables in
constructing the Jacobians. Therefore, at the continuum level, basic
theorems can be trivially applied to our formulation. To avoid
coupling through derivatives between the Jacobian and evolution
equations, which would require a more careful mathematical analysis,
it was crucial that we could replace first derivatives of the metric
using the reduction constraints. This works because the expansion
contains at most one derivative of the metric. Since this fact remains
true even in the absence of spherical symmetry we hope, eventually, to
generalize the DF-excision strategy to the full~$3+1$ setting. For now
it is unclear whether or not this will pan out, since there is a
qualitative difference between 1d and 3d excision that cannot be
overlooked. But if successful the generalization would provide an
improved moving excision strategy for binary black holes within a
pseudospectral code.

In order to perform our numerical tests, appropriate boundary
conditions at the outer boundary must be provided. At present, we do
not yet have outer boundary conditions in the code for the DF
projects. To overcome this problem at the outer boundary, we ensure
that the Jacobian transitions into the identity Jacobian at the
outermost subpatch. This is achieved by using a low order polynomial
transition function, which works remarkably well in practice when the
transition is placed at the two ends of the penultimate subpatch, and
we expect that various alternative configurations would behave
similarly. A desirable alternative would be to implement outer
boundary conditions in the code that take care of the full DF
infrastructure, including the management of two time coordinates. Work
on this will be reported on in the near future. An immediate goal is
to use the methods developed here to study systematically, in the
spherical context, the transition from the linear regime we studied
in~\cite{BhaHilNay20} to the case with arbitrary non-linear
perturbations.

\acknowledgements 

We are grateful to Thanasis Giannakopoulos and Isabel Su\'arez
Fern\'andez and for helpful discussions and feedback on the
manuscript. MKB and KRN acknowledges support from the Ministry of
Human Resource Development (MHRD), India, IISER Kolkata and the Center
of Excellence in Space Sciences (CESSI), India, the Newton-Bhaba
partnership between LIGO India and the University of Southampton, the
Navajbai Ratan Tata Trust grant and the Visitors' Programme at the
Inter-University Centre for Astronomy and Astrophysics (IUCAA),
Pune. CESSI, a multi-institutional Center of Excellence established at
IISER Kolkata is funded by the MHRD under the Frontier Areas of
Science and Technology (FAST) scheme.  DH gratefully acknowledges
support offered by IUCAA, Pune, where part of this work was
completed. The work was partially supported by the FCT (Portugal) IF
Program~IF/00577/2015, Project~No.~UIDB/00099/2020 and
PTDC/MAT-APL/30043/2017.

\normalem
\bibliography{DFex}

\end{document}